\documentclass[aps,prb,twocolumn]{revtex4-1}
\usepackage{amsmath}
\usepackage{amssymb}
\usepackage{enumitem}
\usepackage{physics}
\usepackage[dvipsnames]{xcolor}
\usepackage[T1]{fontenc}
\usepackage{graphicx}% Include figure files
\usepackage{dcolumn}% Align table columns on decimal point
\usepackage{bm}% bold math
\usepackage[retainorgcmds]{IEEEtrantools}
\usepackage{xcolor}
\usepackage[colorlinks=true,linkcolor=blue,citecolor=blue,urlcolor=blue]{hyperref}
\usepackage{ulem}
\usepackage{bbold}
\usepackage{mathtools}

\newcommand{\hide}[1]{}

\newcommand{\SEx}{\Sigma^{\mathrm{x}}}
\newcommand{\SEc}{\Sigma^{\mathrm{c}}}

\newcommand{\scrc}{W^{c}}

\newcommand{\irrchi}{\chi^0}

\newcommand{\G}{\mathbf{G}}
\newcommand{\q}{\mathbf{q}}
\newcommand{\vc}{\mathbf{v}}

\renewcommand{\k}{\mathbf{k}}
\newcommand{\dm}{\rho}
\newcommand{\inveps}{\epsilon^{-1}}

\newcommand{\eps}{\epsilon}

\newcommand{\vcoul}{v}

\newcommand{\bfunc}{\bar{f}}
\newcommand{\bbfunc}{\bar{\bar{f}}}

\newcommand{\editor}[2]{%
  \expandafter\newcommand\csname #1note\endcsname[1]{%
    \textcolor{#2}{(\textbf{#1:} \textit{##1})}}%
  \expandafter\newcommand\csname #1\endcsname[1]{%
    \textcolor{#2}{##1}}%
  \expandafter\newcommand\csname #1cancel\endcsname[1]{%
    \textcolor{#2}{\sout{##1}}}%
  \expandafter\newcommand\csname #1change\endcsname[2]{%
    \textcolor{#2}{\sout{##1} ##2}}%
  \newenvironment{#1text}{\color{#2}}{\color{black}}
}
\definecolor{Blu}{rgb}{0.00,0.00,1.00}
\definecolor{Red}{rgb}{1.00,0.00,0.00}
\definecolor{Orange}{rgb}{1.00,0.45,0.00}
\definecolor{tangerine}{rgb}{0.944,0.522,0}
\definecolor{brown}{rgb}{0.633,0.156,0.156}
\definecolor{lime}{rgb}{0.5,1.0,0.0313}
\definecolor{limedark}{rgb}{0.333, 0.666, 0.020}
\definecolor{applegreen}{rgb}{0.55, 0.71, 0.0}
\definecolor{green1}{rgb}{0.0, 0.5, 0.0}
\definecolor{green2}{rgb}{0.25, 0.5, 0.016}
\definecolor{BluBondi}{rgb}{0.00,0.58,0.71}

\editor{DV}{BluBondi}
\editor{AF}{tangerine}
\editor{PDA}{Orange}
\editor{AG}{limedark}
\renewcommand{\emph}{\textit}

\begin{document}

\title{Efficient GW calculations in two dimensional materials through a stochastic integration of the screened potential}
\author{Alberto Guandalini$^{*1}$, Pino D'Amico$^{1}$, Andrea Ferretti$^{1}$, Daniele Varsano$^{1}$}

\affiliation{$^1$S3 Centre, Istituto Nanoscienze, CNR, Via Campi 213/a, Modena (Italy)}
%\author{Alberto \surname{Guandalini}}
%\author{Pino \surname{D'Amico}}
%\author{Andrea \surname{Ferretti}}
%\author{Daniele \surname{Varsano}}

\date{ \today}

\begin{abstract}
Many-body perturbation theory methods, such as the $G_0W_0$  approximation, are able to accurately predict quasiparticle (QP) properties of several classes of materials.
However, the calculation of the QP band structure of two-dimensional (2D) semiconductors is known to require a very dense BZ sampling, due to the sharp $q$-dependence of the dielectric matrix in the long-wavelength limit ($\mathbf{q} \to 0$).
%
% AF: prev version
%due to the sharp $q$-dependence of the dielectric matrix in the long-wavelength limit ($q \approx 0$), it is known that 
%the calculation of the QP band structure of two-dimensional (2D) semiconductors requires a very dense BZ sampling. 
%
%to properly converge the integrals over the BZ, e.g., in the calculation of the correlation part of the self-energy. This  is  due to the sharp $q$-dependence of the dielectric matrix in the long-wavelength limit ($q \approx 0$)
In this work, we show how the convergence of the QP corrections of 2D semiconductors with respect to the BZ sampling can be drastically improved, by combining a Monte Carlo integration with an interpolation scheme able to represent the screened potential between the calculated grid points.
The method has been validated by computing the  band gap of three different prototype monolayer materials: a transition metal dichalcogenide (MoS$_2$), a wide band gap insulator (hBN) and an anisotropic semiconductor (phosphorene).
The proposed scheme shows that the convergence of the gap for these three materials up to 50meV 
%\PDAnote{io toglierei le parentesi ed anche "say", così esplicitiamo senza ambiguità quello che facciamo} 
%\PDAnote{scriverei "is" al posto di "can be" altrimenti sembra che ci possa essere anche l'eventualità che non succeda, mentre i dati mostrano che è così} 
is achieved by using $\k$-point grids comparable to those needed by DFT calculations, while keeping the grid uniform.
%, \AF{and orders of magnitude coarser wrt calculations done without any accelerator}. \AFnote{questo so che e' un punto controverso, ma non ci starebbe male...}
%\DVnote{considerando che andra in mano a qualche acceleratore (Deslippe, Thygesen) mi sembra azzardato.} \AFnote{ok}

%smaller than the ones needed in standard techniques }by a factor 80, 40 and 20 respectively.
%
\end{abstract}

\maketitle
%========================
%\section{Introduction}
%========================
The GW approximation~\cite{Hedin_1965,Strinati_1982,Hybertsen_1986,Godby_1988} is a well-established method for first principle calculations of electronic excitations of materials\cite{reining2018gw,Golze_2019,Marzari2021NatMaterials}. 
It provides access to quasi-particle 
%\PDAchange{energies and bands} 
energy bands as measured in ARPES experiments,~\cite{Damascelli2003RMP}
%\cite{Plummer_1982,Himpsel_1983,Kevan_1992},
%direct \cite{Plummer_1982,Himpsel_1983,Kevan_1992} or inverse} 
%\cite{Dose_1985,Smith_1988,Fuggle_1992} photoemission spectroscopy, being its theoretical counterpart, i.e., a theoretical spectroscopy method.
satellites,~\cite{Guzzo2011PRL,Caruso2018PRB} lifetimes,~\cite{Marini2002PRB} and spectral functions~\cite{Bechstedt1994PRB,Gatti2020PNAS}. 
Since its development, the GW approximation has been applied to a large class of systems ranging from bulk crystals to nanostructures and molecules~\cite{reining2018gw,Golze_2019}.
During the last decades, since the isolation and characterization of graphene~\cite{geim2010rise}, large attention has been devoted to the study of 2D materials, due to  their remarkable electronic and optical properties~\cite{Ferrari_2015,Bhimanapati_2015}.
Since then, the GW approximation has been extensively applied to predict quasi-particle properties of these materials~\cite{trevisanutto2008ab,haastrup2018computational,rasmussen2021towards,qian2014quantum,varsano2020monolayer,qiu2016screening,Qiu_2017,DaJornada_2020,Yabei_2018,Zhang_2019,Qiu_2020}. 
%\cite{Gao_2017,Smart_2018}

Often, 2D systems are treated using plane waves within the supercell approach, in which an amount of vacuum is added in the non periodic direction in order to remove spurious interactions among replicas.
In principle, accurate GW calculations require the inclusion of a very large vacuum extension due to the long-range nature of the Coulomb interaction.
This difficulty has been mitigated 
%\PDAchange{by making use of}
using truncated Coulomb potentials~\cite{Beigi_2006,Rozzi_2006}
%\PDAchange{, allowing one}
that allows one
%feature can be in part overcome truncating the Coulomb interaction
%at a given distance from the system
%In this way, 
to obtain converged results considering
%can be obtained with 
%$\approx 10-20$ \AA \ of vacuum interlayer distance.
manageable interlayer distances (e.g. in the range of 10-20\AA).
Furthermore, characteristic properties of 2D screening, such as the dielectric function %\PDAchange{that approaches}
approaching unity in the long-wavelength limit (see below), are correctly reproduced in the supercell approach only if the potential is appropriately truncated\cite{Cudazzo_2011,Huser_2013}.
%The specificity of 2D systems, in particular the behaviour of the screened potential
However, once the Coulomb potential is truncated, the resulting sharp behavior of the screened potential can make 
%the use of standard techniques for 
the integration over the Brillouin zone (BZ) rather inefficient
~\cite{qiu2016screening,Huser_2013}.
Thus very large $k$-point grids are needed to obtain converged results, making the computation of quasiparticle (QP) properties within the GW method for 2D systems computationally expensive\cite{Thygesen_2017}.

\begin{figure}%[hbtp]
\includegraphics[width=0.5 \textwidth]{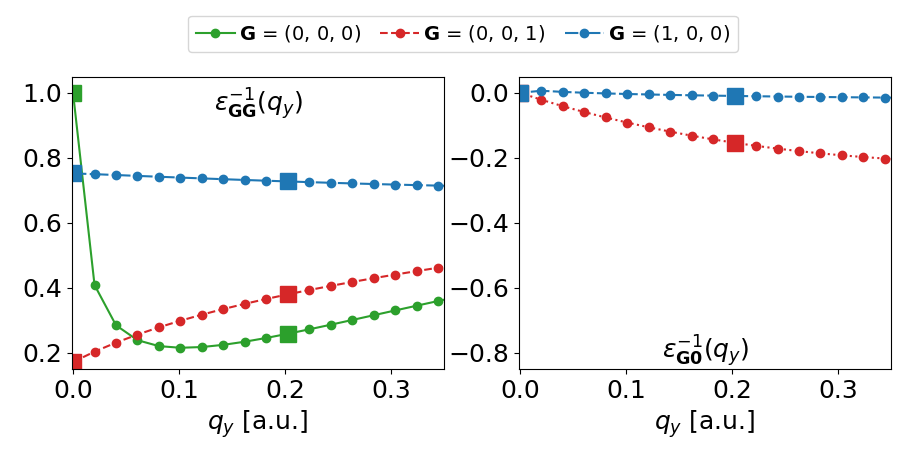}
\caption{Selected elements of the real part of the static inverse dielectric matrix of MoS$_2$. Diagonal and wing elements elements are plotted in the left and right panel respectively. Squares indicate values obtained with a $6\times 6 \times 1$ grid, while dots with a $60\times 60 \times 1$ grid.}
\label{Fig_epsm1}
\end{figure}

%While cutting of the potential is mandatory in order to obtain converged results with respect to the amount of vacuum, at the same time QP properties becomes very challenging to converge with respect to the k-grid mesh.
%\AG{It is instructive to understand here in more detail the 2D screening features which cause the difficulties of integration over the BZ previously emphasized.}
More in details,
in a plane-wave basis set description, the screening properties are described by the matrix elements of the Fourier transform of the inverse dielectric function $\epsilon^{-1}_{\mathbf{G}\mathbf{G}'}(\mathbf{q})$, where $\G$ is a reciprocal lattice vector and $\q$ a reciprocal vector of the first BZ. 
%As an example, some selected matrix elements of monolayer MoS$_2$ are shown in Fig. \ref{Fig_epsm1}.
 In 2D systems, as already pointed out in the literature\cite{Cudazzo_2011,huser2013dielectric,Qiu_2016,Rasmussen_2016,daJornada_2017,Chernikov_2014}, the head  [$\mathbf{G} = \mathbf{G}' = (0,0,0)$] of the dielectric matrix sharply approaches unity in the long wavelength limit (Fig.~\ref{Fig_epsm1}, left panel), and it is clear that with coarse meshes 
 %\PDAchange{are not able to follow}
 it is not possible to correctly reproduce such limit with a regular discretization procedure.
 In addition, the first matrix elements associated with lattice vectors along the confined direction ($\mathbf{G}_{\perp}$) show a dispersion in the long-wavelength limit, differently from the matrix elements with lattice vectors oriented in the periodic directions ($\mathbf{G}_{\parallel}$) which are approximately constant with respect to $|\q|$ (see Fig.~\ref{Fig_epsm1}).
 This trend originates from the fact that $\text{min} |\mathbf{G}_{\perp}|$ is significantly smaller than $\text{min} |\mathbf{G}_{\parallel}|$, due to the amount of vacuum added in the perpendicular direction.
 %As an example, for MoS2 with $10$\AA{} of vacuum distance between replicas (the system considered in Fig.~\ref{Fig_epsm1}), we have $ \min|\mathbf{G}_{\parallel}|=1.21$ a.u., while  $ \min|\mathbf{G}_{\perp}|= 0.33$ a.u.,
 %
%only slightly larger than the $\mathbf{k}$-point discretization step ($|\Delta \mathbf{q}|= 0.20$ a.u.)
% This trend originates from the fact that the electron density of 2D systems is 
 %is nearly homogeneous along the periodic directions, while
 %highly inhomogeneous (at least) along the confined direction.
%\AGnote{Ho sostituito la motivazione del perché gli elementi di matrice corrispondenti ai G ortogonali sono più dispersivi, seguendo il ragionamento di Mauri.}\\
%\DVnote{vedi comment}\AGnote{Answered}
%\AFnote{anche a me pare un dettaglio eccessivo. Peraltro, se proprio lo vogliamo mettere (io toglierei cmq), userei almeno un vuoto di 15 o 20 \AA{}, che almeno evidenzia l'effetto... cosi' mi pare anche poco convicente.}\DVnote{diamo il $\Delta$q senza menzionare il k point sampling. DA DISCUTERE}
%\AFnote{PS: ho messo tutti gli uguali sui |G| perche' sono valori precisi.}
Furthermore, we note that in 2D systems the long-wavelength limit of the wing matrix of the dielectric matrix, $\epsilon^{-1}_{\mathbf{G}\mathbf{0}}$, goes to zero as $|\mathbf{q} \to 0|$ (Fig.~\ref{Fig_epsm1} right panel)
leading to possible numerical 
%\PDAchange{issues}
instabilities when these terms are multiplied by the diverging Coulomb potential.
%As these terms are multiplied by a divergent Coulomb interaction for the calculation of the screened potential (see below), numerical problems may arise.
All these features of the dielectric matrix 
contribute to slow the convergence of the QP properties with respect to the number of sampling points in the BZ, usually discretized on a uniform grid.
%\AG{In conventional 3D semiconductors, where all the reciprocal lattice vectors are oriented along periodic directions ($\G = \mathbf{G}_{\parallel}$), loose meshes (e.g. squares in Fig.~\ref{Fig_epsm1}) are instead enough to properly describe the behavior of the inverse dielectric matrix as a function of momentum transfer.}
%\DVnote{vedi comment}\AGnote{Answered}

In the last years, different strategies have been proposed to accelerate the convergence of GW results for 2D systems with respect to the number of $\mathbf{k}$-point sampling.
Rasmussen et al.\cite{Rasmussen_2016} proposed an analytic model for the long-wavelength limit of the head of the inverse dielectric function $\epsilon^{-1}_{\mathbf{0}\mathbf{0}}$.
This model has been used to integrate the screened potential in a small region around the $\Gamma$ point, thus reducing the size of the $\k$-point mesh needed to converge the quasiparticle gap.
However, denser meshes with respect to DFT are still required (e.g. for the band gap of MoS$_2$ converged results within 50meV were reported\cite{Rasmussen_2016} using  $18 \times 18 \times 1$ grids)
%(converged results with the $\approx 18 \times 18 \times 1$ for the band gap of MoS$_2$), 
as the analytic model is applied only to the $\G = \G' = 0$ matrix element. 

Da Jornada et al.~\citep{daJornada_2017} proposed instead a fully numerical approach, where a nonuniform $\mathbf{q}$-sampling is used to increase the sampling close to the $\Gamma$ point.
This approach has been applied
not only to the $\G = \G' = 0$ element of the dielectric matrix, but to a submatrix 
%composed by some
($\mathbf{G}_{\perp}$, $\mathbf{G}'_{\perp}$) such that $|\mathbf{G}_{\perp}|$,$|\mathbf{G}_{\perp}'| < \min|\mathbf{G}_{\parallel}|$. %\DVcancel{, where $\mathbf{G}_{\perp}$ and $\mathbf{G}_{\parallel}$ are out-of-plane and in-plane $G$ vectors.}\DVnote{già definiti prima}
In Xia et al.~\citep{Xia_2020}, 
the two previous strategies are combined by performing a non-uniform sub-sampling of the Brillouin zone around $\Gamma$ followed by a non-linear fitting procedure to model the $\mathbf{q}$-dependence of the self-energy terms (both exchange and correlation) instead of modeling the behaviour of the dielectric matrix or screened potential elements.

%\DVchange{
%These last two methods  are able to accelerate the convergence of the GW gap with respect to the size of the $k$-point grid providing, 
%\AG{for instance for the 2D benchmark systems they considered (MoSe$_2$ bilayer and MoS$_2$ monolayer respectively)}, reliable results using meshes similar to the ones needed to converge the DFT ground state calculation.}
{The last two methods showed that meshes of size similar to those needed to converge the DFT ground state calculation were enough to obtain reliable results as demonstrated for the gap of MoSe$_2$ bilayer and MoS$_2$ monolayer.}
%In fact, to sample the q-space in $N$ additional points around $\Gamma$ requires to sample the k space in $N$ shifted grids. This can be problematic if the method is applied to large-scale systems.
%
However, both methods rely on a nonuniform sampling, which add a convergence parameter to be managed (the number of sub-sampling points).
%In addition, to add $N$ points around $\Gamma$ in the q sampling requires to compute the KS states on $N$ slightly-shifted \AGcancel{uniform} k grids, adding a computational step in the calculation.\DVnote{see comment}\AGnote{Ho riscritto questa parte}
%\AFcancel{The computational cost due to the additional sampling is heavily counterbalanced by the reduction of the grid size, which result in a global acceleration process. Still, algorithms which does not rely on additional samplings of the BZ would be preferable.}
%However, for large systems, the use of $N$ uniform grids can be problematic.
%In fact, memory is often a critical issue in modern HPC architectures, which heavily rely on GPUs with low dedicated memory[REFs].\DVnote{see comment}
Moreover, the region around $\Gamma$ in which the additional sampling is performed (and consequently the nonlinear fitting in Xia et al.) depends on the size of the uniform grid.
This may cause inconsistency problems when comparing results from different grids, e.g., in a convergence set of calculations, as increasing the grid the region around $\Gamma$ becomes smaller and smaller.
%\\
%A method which is able to improve the integration accuracy on a region that does not depend on the uniform grid discretization is thus preferable.
%Finally, we note the method of Xia et al. relies on a fitting of the self energy instead of the inverse dielectric function (or alternatively the screened potential), thus it cannot be easily extended to other kind of many-body perturbation theory calculations, e.g., to solve the Bethe-Salpeter equation.

Motivated by these works, we show that the convergence of QP properties of 2D semiconductors with respect to the number of $\mathbf{k}$-points in the BZ sampling can be accelerated, at the same level of previous methods found in the literature\cite{daJornada_2017,Xia_2020}, by combining the Monte Carlo integration techniques with an interpolation scheme of the screened potential.
Unlike the methods described above, the proposed method allows one to accelerate the convergence of the QP properties overcoming the need to rely on a nonuniform sampling.
%thus avoiding the need to compute and store the KS states on $N$ uniform grids.
%\DVnote{see comment}\AGnote{Done}
%
In addition, the same integration procedure (see below) is applied to the full BZ, thereby avoiding the need to treat the $\Gamma$ region differently from the remaining part of the BZ.
The proposed method has been implemented in the Yambo package~\citep{yambo_2009,yambo_2019}.

The work is organized as follows: In Sec.~\ref{Sec_methods} we present the main ideas of the proposed method and its implementation. In Sec.~\ref{results}, we show the performance of the method for three prototype 2D semiconductors:
a transition metal dichalcogenide (MoS$_2$), a wide band gap insulator (hBN), and an anisotropic semiconductor (phosphorene). In Sec.~\ref{compdet} we provide the computational details and in Sec.\ref{concl} we draw the conclusions.
In Appendix~\ref{Appx1} we provide the derivation of the expression used in the proposed method for the head of the screened potential matrix.

%\section{Some notes on stuff that should be written}
%``'$\Sigma_{n\mathbf{k}}(\mathbf{q})$ is the contribution to the GW self-energy for state nk from point q in the BZ, Omega is the volume of the BZ and fq is the appropriate weight.''\\
%N.B. cite 16-21 of da Journada when you talk about sharp q-dependence of the dielectric function. There is also some stuff in 24-25 of thyghesen.\\
%``The large cell volume translates into the need to include a large number of electronic states in GW calculations'' (this will justify the use of terminators).\\
%``This method can be readidly incorporated into several ab initio packages thta compute electronic and optical properties through many-electron perturbation theory methods.''\\
%``Da Journada et al introduce the terms necks for the (Gz,Gz) elements of the dielectric function."\\
%``We start by taking the continuous limit and rediscretizing''

%========================
\section{Methods}\label{Sec_methods}
%========================

Within many-body perturbation theory, quasiparticle energies are usually calculated either by solving numerically the QP equation:
\begin{equation}
\varepsilon_{n\k}^{\text{QP}} =  \varepsilon_{n\k}^{\text{KS}}+ \mel{n\k}{\Sigma(\varepsilon_{n\k}^{\text{QP}}) - v_{xc}^{\text{KS}}}{n\k} \ ,
\end{equation}
where $\{n\k\}$ are the KS wavefunctions and  $v^{\text{KS}}_{xc}$ is the exchange-correlation potential, 
or by linearizing the equation at the first order:
\begin{equation}
\varepsilon_{n\k}^{\text{QP}} =  \varepsilon_{n\k}^{\text{KS}}+ Z_{n \bf k}\mel{n\k}{\Sigma(\varepsilon_{n\k}^{\text{KS}}) - v_{xc}^{\text{KS}}}{n\k},
\end{equation}
%
%In the latter expression the renormalization factor $Z_{n \bf k}$ is computed from the first term of the Taylor expansion of the self energy, $\Sigma$: 
where the renormalization factor $Z_{n \bf k}$ is defined as:
\begin{equation}
   Z_{n\mathbf{k}} = \left[\left. 1-\langle n {\bf k}|\frac{\partial\Sigma(\omega)}{\partial \omega}| n {\bf k}  \rangle\right|_{\,\omega=\epsilon_{n {\bf k}}^{\text{KS}}} \right]^{-1} .
   \label{qp_z}
\end{equation}

To obtain the QP correction of a single-particle state $\ket{n\mathbf{k}}$ in the GW approximation, we need to evaluate the diagonal matrix element of the self-energy,
%\AF{
\begin{equation}
    \Sigma^{GW}(\omega) = 
     -\int_{-\infty}^{+\infty} 
     \frac{d\omega'}{2 \pi i}  e^{i \omega' 0^+}, G(\omega+\omega') W(\omega'),
    \label{eq:GW}
\end{equation}
where the screened interaction $W$ is obtained from the expression $W(\omega)=v + v \chi(\omega)v = \inveps(\omega)v $, with the reducible polarizability $\chi(\omega)$ treated at the RPA level.
%}
%
The self energy can be split into the exchange (x) and correlation (c) parts as
\begin{equation}\label{Eq_S}
\mel{n\k}{\Sigma(\omega)}{n\k} \equiv \SEx_{n\k}+\SEc_{n\k}(\omega).
\end{equation}
%Notably, both terms of the self-energy involve a $\mathbf{q}$-integration over the BZ.
%\AG{By discretizing the $\mathbf{q}$-integral into a finite sum over a 2D Monkhorst-Pack grid\citep{Monkhorst_1976}, the x self-energy is written as:}{By discretizing the BZ with a 2D Monkhorst-Pack grid\citep{Monkhorst_1976},}
Notably, both terms of the self-energy involve an integration over the momentum transfer $\q$. If we discretize the BZ with a 2D uniform $\k$-grid (centered at $\Gamma$), following the Monkhorst-Pack scheme\citep{Monkhorst_1976}, the momentum transfer $\q$ is discretized with the same uniform grid, and the $\q$ integrals can be evaluated as finite sums.
Thus, the x self-energy is written as: 
\begin{equation}\label{Eq_Sx}
\SEx_{n\k} = -\frac{1}{N_{q}\Omega}\sum\limits_{v,\q}
\sum\limits_{\G} |\dm_{nv}(\k,\q,\G)|^2 \vcoul_{\G}(\q) \ ,
\end{equation}
where $\Omega$ is the volume of the unit cell in real space, $v$ labels the occupied bands, the $\dm_{nm}$ matrix elements are defined as $\dm_{nm}(\mathbf{k},\mathbf{q},\mathbf{G})=  \mel{n\mathbf{k}}{e^{i(\mathbf{q}+\mathbf{G}) \cdot \mathbf{r}}}{m\mathbf{k-q}}$, and  $N_{q}$ is the number of points of the $q$ grid.
In order to eliminate periodic image interactions for a 2D system, we take the Coulomb potential in Eq.~\eqref{Eq_Sx} as a truncated potential in a slab geometry.
Its Fourier transform reads~\cite{Beigi_2006, Rozzi_2006}: 
\begin{multline}\label{Eq_cutoff_q}
\vcoul_{\G}(\q) = \frac{4\pi}{|\q+\G|^2} \\
%\times \left[ 1 - e^{-|\q_{xy}+\G_{xy}|L/2}\cos[(q_z+G_z) L/2] \right] \ ,
\times \left[ 1 -e^{-|\mathbf{q}_{\parallel}+\mathbf{G}_{\parallel}|L/2}\cos[(q_z+G_z) L/2] \right] \ ,
\end{multline}
where $L$ is the length of the cell in the non-periodic $z$ direction.
As the q-grid is 2D, we have $q_z = 0$.

Nevertheless, 
Eq.~\eqref{Eq_Sx} cannot be directly applied due to the divergence of the Coulomb interaction at $\mathbf{G}=\mathbf{q}=0$.
There are several approaches to treat such divergence~\cite{Gygi_1986,carrier2007general,deslippe2012berkeleygw,pulci1998ab,yambo_2009}.
Among these, we select the v-average (v-av) method (called random integration method and described in Ref.~\onlinecite{yambo_2009}).
In this method,  it is assumed that the matrix elements $\dm_{nm}(\mathbf{k}, \mathbf{q}, \mathbf{G})$ are smooth with respect to $\q$, and Eq.~\eqref{Eq_Sx} is discretized in the following way
\begin{equation}\label{Eq_Sx_av}
\SEx_{n\k} = -\frac{1}{N_{q}\Omega}\sum\limits_{v,\q}
\sum\limits_{\G} |\dm_{nv}(\k,\q,\G)|^2 \bar{\vcoul}_{\G}(\q) \ ,
\end{equation}
where $\bar{\vcoul}$ is the average of the Coulomb interaction within a region of the BZ centred around $\q$ of the Monkhorst-Pack grid:
\begin{equation}\label{Eq_v_av}
\bar{\vcoul}_{\G}(\q) = \frac{1}{D_{\bf \Gamma}}\int\limits_{D_{\bf \Gamma}}\frac{d\q'}{(2\pi)^2}\vcoul_{\G}(\q+\q') \ .
\end{equation}
$D_{\Gamma}$ is the small area of the Monkhorst-Pack grid centered around $\Gamma$
(red area in Fig.~\ref{fig_BZ}).
%\DVnote{check: direi che viene fatto per ogni q, non solo a Gamma, ho sostiuito nell'equazione e testo $D_\Gamma$ con $D_q$}.
The integrals in Eq.~\eqref{Eq_v_av} are evaluated via a 2D Monte Carlo technique.
%\AFcancel{The regularization given by Eq.~\eqref{Eq_v_av} remedies the divergence problem of the Coulomb interaction.}
%\AFnote{lo abbiamo gia' detto sopra, direi.}

In addition,
%\PDAnote{ volevi dire "inserting ... in " ?}
Eq.~\eqref{Eq_Sx_av}, as compared with Eq.~\eqref{Eq_Sx},  leads to a faster convergence of the the exchange self-energy with respect to $N_q$, since Eq.~\eqref{Eq_Sx_av} takes into account the variation of the bare potential within the region of the BZ centred around each $\bf q$ point and it has been extensively applied also to three dimensional semiconductors~\cite{rangel2020reproducibility}.
For practical purposes, it is sufficient to evaluate the averages up to a threshold $|\G|^2/2 < E_{\mathrm{cut}}^{\text{v-av}}$, 
since the Coulomb interaction becomes a smooth function of $\q$ at large $|\G|$.
%\AFnote{invece di "lim", vogliamo usare v-av ?}
%\DVnote{"lim" non piace neanche a me $\G^{\text{thr}}$?}

We now consider the correlation part of the self-energy, that is the most problematic term for 2D semiconductors.
Within the plasmon-pole approximation (PPA) (we here adopt the Godby-Needs formulation~\cite{Godby_1989}) 
the correlation part of the screened Coulomb potential, $W^c(\omega)=W(\omega)-v$, 
%=v\chi(\omega)v
is written as:
\begin{equation}
\label{Eq_W}
   \scrc_{\G\G'}(\q,\omega) = \frac{2R_{\G\G'}(\q)\Omega_{\G\G'}(\q)}{\omega^2-[\Omega_{\G\G'}(\q)-i\eta]^2},
\end{equation}
where the limit $\eta\to 0^+$ is implicitly assumed.
Then, $\SEc$ can be expressed as:
\begin{equation}\label{Eq_Sc}
\SEc_{n\k}(\omega) = \frac{1}{N_q \Omega} \sum\limits_{\G,\G',\q} g_{\G\G'}^{n\k}(\q,\omega)\, \scrc_{\G\G'}(\q)   \ ,
\end{equation}
where the matrix elements $g_{\G\G'}^{n\k}$ are defined as:
% \DVnote{see comment}\AGnote{Answered}
%
\begin{multline}\label{Eq_g}
g_{\G\G'}^{n\k}(\q,\omega) = %\lim\limits_{\eta \to 0^+}
-\frac{1}{2}\sum\limits_{m}
\\ \times
%\\
%\frac{1}{2}\sum\limits_{m}
\frac{\dm_{nm}(\k,\q,\G)\,\Omega_{\G\G'}(\q)\,\dm_{nm}^*(\k,\q,\G')}{\omega -\varepsilon_{m\k-\q}^{\text{KS}} +[\Omega_{\G\G'}(\q)-i\eta]\,\text{sgn}(\mu-\varepsilon_{m\k-\q}^{\text{KS}})},
\end{multline}
%
%\AF{where $\Omega_{\G\G'}(\q)$ are the poles of the response function obtained here within the PPA},
%\DVnote{see comment}\AGnote{Answered}
$\mu$ being the chemical potential.
%\AFnote{Il fattore 1/2 della (11) da dove viene ?}
%\AF{In particular}, $g_{\G\G'}^{n\k}$ are smooth functions of $\q$ for $\omega$ \AF{far from} $\Omega_{\G\G'}$,
%$\omega \ll \Omega_{\G\G'}$, 
%where the PPA is justified. 
%
In Eq.~\eqref{Eq_Sc}
$\scrc_{\G\G'}(\q)$ is the static component of the screened Coulomb interaction.
%
%\begin{equation}\label{Eq_W}
%\scrc_{\G\G'}(\q) = \sqrt{\vcoul_{\G}(\q)} \tilde \eps^{-1}_{\G\G'}(\q)\sqrt{\vcoul_{\G'}(\q)}.
%\end{equation}
%In Eq.~\eqref{Eq_W}, $\tilde \eps^{-1}_{\G\G'}$ is the correlation part of the static symmetrized \DVnote{symmetric?} inverse dielectric function
%\begin{equation}\label{Eq_inv_eps}
%\tilde \eps^{-1}_{\G\G'}(\q) =  \sqrt{\vcoul_{\G}(\q)}\chi_{\G\G'}(\q)\sqrt{\vcoul_{\G'}(\q)},
%\end{equation}
%where $\chi_{\G\G'}$ is the reducible density-density response function in the RPA approximation.
%
In particular, $g_{\G\G'}^{n\k}$ are smooth functions of $\q$ for $\omega$ far from $\Omega_{\G\G'}$,
%$\omega \ll \Omega_{\G\G'}$, 
where the PPA is justified.

For small $\mathbf{G}$ vectors, the correlation part of the screened potential, $\scrc_{\G\G'}(\q)$ shows a sharp $\q$ dependence due to the behaviour of both the bare interaction $\vcoul_{\G}(\bf q)$ and the inverse dielectric function $\inveps_{\G\G'}(\bf q)$, as shown in Fig.~\ref{Fig_epsm1}.
For this reason, following a similar procedure already applied to $\SEx_{n\k}$, we discretize Eq.~\eqref{Eq_Sc}  as
\begin{equation}\label{Eq_Sc_av}
\SEc_{n\k}(\omega) = \frac{1}{N_q \Omega} \sum\limits_{\G,\G',\q} g_{\G\G'}^{n\k}(\q,\omega) \overline{\scrc}_{\G\G'}(\q)   \ ,
\end{equation}
where
\begin{equation}\label{Eq_W_av}
\overline{\scrc}_{\G\G'}(\q) = \frac{1}{D_{\Gamma}}\int\limits_{D_{\Gamma}}\frac{d\q'}{(2\pi)^2}\scrc_{\G\G'}(\q+\q')
\end{equation} 
is the average of the correlation part of the screened potential in the mini-BZ around $\q$ of the Monkhorst-Pack grid.
The evaluation of the correlation part of the self-energy via Eq.~\eqref{Eq_W_av} is referred to in the following as the W-av method.
The integrals in Eq.~\eqref{Eq_W_av} are calculated using a 2D Monte Carlo integration method,
where $\overline{\scrc}_{\G\G'}(\q+\q')$ is evaluated considering typically $\approx 10^{6}$ $\q'$ points in the region around $\Gamma$ (red area of Fig.~\ref{fig_BZ}) making use of an interpolation scheme, that is discussed in details in the next section.
%first, we interpolate $\overline{\scrc}_{\G\G'}(\q+\q')$ between the $\q$ point and its nearest neighbours in the Monkhorst-Pack grid (as explained in the next section);
%second, we evaluate $\overline{\scrc}_{\G\G'}(\q+\q')$ in a set of $\approx 10^{6}$ random points to perform the Monte Carlo integration in Eq. (); finally, the $\overline{\scrc}_{\G\G'}(\q)$ are used to evaluate the c self energy in Eq. ().
In practice, the Monte Carlo average is performed for a limited number of matrix elements of $W$ such that $|\G|^2/2 < E_{cut}^{\mathrm{W-av}}$, %\AFcancel{where $E_{cut}^{av}$ is a given cutoff,} 
i.e. for the matrix elements presenting a sharp behaviour as a function of $\q$, while for the remaining $\mathbf{G}$ vectors %matrix elements such that  $|\G|^2/2 > E_{cut}^{av}$ 
the screening is evaluated on the $\q$ grid determined by the $\k$-point sampling.

%In order to evaluate the screened potential at million of points, $\scrc_{\G\G}(\q+\q')$ is locally approximated with parametric functions with the interpolation algorithm explained in the next section.
Importantly, the W-average correction performed for $|\G|^2/2 <E_{cut}^{\mathrm{W-av}}$ is applied to every $\q$ point of the BZ, at variance with other proposed methods where corrections are applied to the $\q = 0$ term only\cite{Rasmussen_2016,daJornada_2017,Xia_2020}.

%========================
\subsection{Interpolation of the static screening}\label{Sec_int_W}
%========================

\begin{figure}%[hbtp]
\includegraphics[width=0.5 \textwidth]{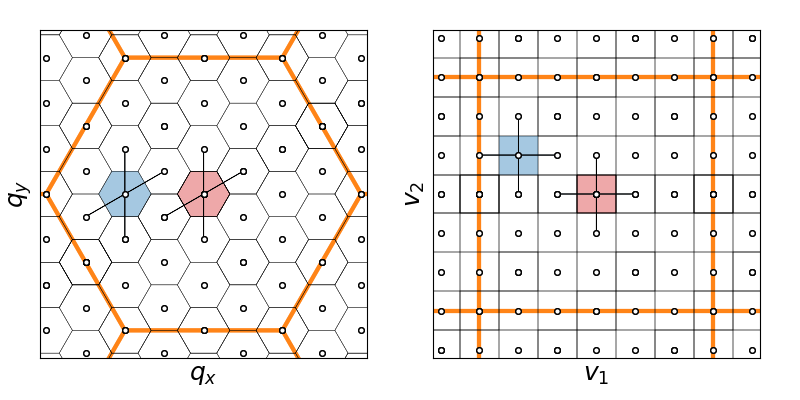}
\caption{Graphical representation of the $6\times 6 \times 1$ Monkhorst-Pack sampling of an hexagonal units cell. In the left panel, the reciprocal space is represented in cartesian coordinates ($\q$), in the right panel in reciprocal lattice coordinates ($\vc$).
A Monkhorst-Pack grid is always rectangular (or squared) in reciprocal lattice coordinates. 
The BZ in both representations are highlighted with thick orange lines.
Smaller hexagons (squares) represent the mini-BZ of the q sampling in cartesian (reciprocal lattice) coordinates. The mini-BZ at $\Gamma$ is highlighted in red, while an example at $\q \neq 0$ ($\vc \neq 0$) in blue. Black lines connect the q point with its nearest neighbours in reciprocal lattice coordinates, which are used for the interpolation.}
\label{fig_BZ}
\end{figure}

In this Section, we describe a procedure to interpolate the correlation part of the static screened potential $\scrc_{\G\G'}(\q)$ as a function of $\q$, for the computation of the average $\overline{\scrc}_{\G\G'}$ according to Eq.~\eqref{Eq_W_av}.
%As a first step, we manipulate the screening such that it can be interpolated easier.
The head of the screened potential, $\scrc_{00}$, can be exactly written as:  
\begin{equation}\label{Eq_Dec_W}
\scrc_{00}(\q) = \frac{\vcoul_{0}(\q)f(\q)\vcoul_{0}(\q)}{1-\vcoul_{0}(\q)f(\q)} \ ,
\end{equation}
 where $f(\q)$ is an auxiliary function defined in  Appendix ~\ref{Appx1}.
The expression of Eq.~\eqref{Eq_Dec_W} suggests that it is possible to use $f(\q)$ for the interpolation of $\scrc_{00}(\q)$.
In fact, while $\scrc_{00}(\q)$ shows a sharp behavour as a function of $\q$, the function $f(\q)$ is smoother, as it resembles the irreducible response function $\irrchi_{00}(\q)$ plus some corrections due to local-field contributions.
In fact, $f(\q) = \irrchi_{00}(\q)$ if local field effects are neglected.
%\DVnote{why $\approx$?}\AFnote{anche a me sembrerebbe che nell'ipotesi di no-local field, si possa usagre l'=... o sbaglio?}
Guided by this argument, we propose to represent the matrix elements of the static screening as
\begin{multline}\label{Eq_W_fit}
\scrc_{\G\G'}(\q+\q') = \\
\frac{\vcoul_{\G}(\q+\q')f_{\G\G'}(\q+\q')\vcoul_{\G'}(\q+\q')}{1-\sqrt{\vcoul_{\G}(\q+\q')}f_{\G\G'}(\q+\q')\sqrt{\vcoul_{\G'}(\q+\q')}} \ ,
\end{multline}
where the Coulomb interaction $\vcoul_{\G}$ is given by Eq.~\eqref{Eq_cutoff_q} and $f_{\G\G'}$ is an auxiliary function.
We note that Eq.~\eqref{Eq_W_fit} is the simplest generalization of Eq. \eqref{Eq_Dec_W} for the case $\G \neq 0$, $\G' \neq 0$.
We remind that in our notation $\q$ is a point of the Monkhorst-Pack grid, while $\q'$ belongs to $D_{\Gamma}$ (red region in Fig.~\ref{fig_BZ}).
By inverting Eq.~\eqref{Eq_W_fit} on the $\mathbf{q}$-points of the mesh we have: 
\begin{multline}\label{Eq_f_num}
f_{\G\G'}(\q) = \frac{\scrc_{\G\G'}(\q)}{\sqrt{\vcoul_{\G}(\q)}\sqrt{\vcoul_{\G'}(\q)}}\\
\times \left[\scrc_{\G\G}(\q)+ \sqrt{\vcoul_{\G}(\q)}\sqrt{\vcoul_{\G'}(\q)} \right]^{-1} \ .
\end{multline}
In order to compute $f_{\G\G'}(\q+\q')$ without requiring a dense mesh of $\q$ points, the function is numerically determined by interpolating between the given $\q$ point and its four nearest neighbors in reciprocal lattice coordinates $\vc$.
A sketch of the interpolation scheme is shown in Fig.~\ref{fig_BZ}.

The auxiliary function is parametrized as:
\begin{multline}\label{Eq_f_fit}
f_{\G\G'}(\vc+\vc') \equiv f_{\G\G'}(\vc) +  \bfunc_{\G\G'}(\vc)\cdot\vc'\\
+\vc'\cdot\bbfunc_{\G\G'}(\vc)\cdot\vc' ,
\end{multline}
where
\begin{equation}\label{Eq_f_coeff_v}
\bfunc_{\G\G'}(\vc) = \begin{bmatrix}
f^{1}_{\G\G'}(\vc) &
f^{2}_{\G\G'}(\vc)
\end{bmatrix}
\end{equation}
and,
\begin{equation}\label{Eq_f_coeff_m}
\bbfunc_{\G\G'}(\vc) = \begin{bmatrix}
f^{11}_{\G\G'}(\vc) & f^{12}_{\G\G'}(\vc)\\
f^{21}_{\G\G'}(\vc) & f^{22}_{\G\G'}(\vc) 
\end{bmatrix},
\end{equation}
$\vc$ and $\vc'$ being $\q$ and $\q'$ in reciprocal lattice coordinates.
The polynomial dependence of $f_{\G\G'}(\vc+\vc')$ with respect to $\vc'$ is inspired by the Taylor expansion of $f_{\G\G'}$ around $\vc$.
In Eq. \eqref{Eq_f_fit}, there are six coefficients that must be determined.
%where the superscript $1$ and $2$ stands for the first and second lattice vector in the periodic directions.
As there are only four nearest neighbors, thus four conditions to apply, we set for simplicity $f^{12}_{\G\G'}(\vc) = f^{21}_{\G\G'}(\vc) = 0$. This choice corresponds to adopt a bilinear interpolation.

%\AG{
We note that $f_{00}(\q \to 0) $ is the most relevant element in the integration of the self-energy~\cite{Rasmussen_2016}.
%}
For semiconductors, it is possible to exploit the known behaviour  $\lim_{\q \to 0} f_{00} (\q) \propto |\q|^2$  (see Eq. \eqref{usl2003} and Ref. ~\onlinecite{Pick_1970})
%  when $\q \to 0$,
%$f_{00} (\q)$ tends to $0$ as $|\q|^2$[see Eq. \eqref{usl2003} and Ref. ~\citep{Pick_1970}.
%\AG{Exploiting the previous condition}, 
to impose a specific and more accurate functional form to the head ($\G = 0$ and $\G' = 0$) at $\q = 0$.
Following the model for the inverse dielectric function adopted by Ismail-Beigi~\citep{Beigi_2006} we consider 
for $f_{00}(\vc')$ the expression:

%Guided by the modelling of the inverse dielectric function in the work of Beigi~\citep{Beigi_2006}, we propose the following functional form
%
\begin{equation}\label{Eq_f00_fit}
f_{00}(\vc') \equiv \q'\cdot \bar{\bar{f}}_{\mathrm{lim}}\cdot\q'e^{-\sqrt{\alpha^2 v_1^{\prime 2}+\beta^2 v_2^{\prime 2}}} \ .
\end{equation}
%
%\AFnote{tutto funzione di v', ma poi nel tensore mettiamo i q'... si riesce a scriverlo diversamente ?}
%
where $\bar{\bar{f}}_{\mathrm{lim}}$ is a $2\times 2$ tensor which describes the anisotropy of $\irrchi_{00}$ and of $\scrc_{00}$, and $\q' = \q'(\vc')$.
%\AG{
We note that in Eq.~\eqref{Eq_f00_fit} the $\bar{\bar{f}}_{\mathrm{lim}}$ tensor is represented in cartesian coordinates. %\DVcancel{,instead of reciprocal lattice coordinates, as we have done in Eq.s~\eqref{Eq_f_coeff_v} and \eqref{Eq_f_coeff_m}}. 
However, we stress that the representation basis is arbitrary, as the tensorial scalar product does not depend on the coordinate choice. %\DVchange{We believe cartesian coordinates is the natural choice to represent $\bar{\bar{f}}_{\mathrm{lim}}$, as,}{
This choice, differently from the reciprocal lattice unit representation, makes the $\bar{\bar{f}}_{\mathrm{lim}}$ proportional to the identity matrix in the case of isotropic systems.
%\DVcancel{if the system is isotropic, $\bar{\bar{f}}_{\mathrm{lim}}$ is proportional to the identity matrix (not true in reciprocal lattice coordinates).}
%}
%\AG{In Eq.~\eqref{Eq_f00_fit}, we note that $\q'$ implicitly depends on $\vc'$ by a coordinate transformation.}
%We chose to define $\bar{\bar{f}}_{\mathrm{lim}}$ in cartesian coordinates (instead of reciprocal lattice coordinates) 
We can partially account for the anisotropy of the auxiliary function by keeping the diagonal form ($f^{xy}_{\mathrm{lim}} = f^{yx}_{\mathrm{lim}}=0)$ but relaxing the proportionality to the identity matrix ($f^{xx}_{\mathrm{lim}} \neq f^{yy}_{\mathrm{lim}}$ ).

%\PDAnote{Ne parlavo con Alberto ieri in istituto: qui diciamo che l'anisotropia si può parzialmente introdurre considerando una flim diagonale. Torna, però volendo differenziare rispetto al caso precedente (flim proporzionale all'identità nel caso isotropo) potremmo forse scrivere una cosa tipo"... auxiliary function by keeping the diagonal form ($f^{xy}_{\mathrm{lim}} = f^{yx}_{\mathrm{lim}}=0)$ but relaxing the proportionality to the identity matrix ($f^{xx}_{\mathrm{lim}} \neq f^{yy}_{\mathrm{lim}}$ )"}.
%\AGnote{A me va bene}
By substituting Eq.~\eqref{Eq_f00_fit} into Eq.~\eqref{Eq_f_num}, and taking the $|\q| \to 0$ limit along the $x$ and $y$ directions, respectively (the periodic directions), %\AFnote{x e y corrispondono ai due generatori del reticolo ? la mesh q non e' necessariamente orthogonale, mentre lo e' la mesh v... qui dobbiamo usare v' o q' ?}
we have
\begin{equation}\label{Eq_flim_anis}
\begin{cases}
f^{xx}_{\mathrm{lim}} = \frac{\scrc_{00}(q'_x\to 0,q'_y=0)}{(2\pi L)^2}
\\
f^{yy}_{\mathrm{lim}} = \frac{\scrc_{00}(q'_x= 0,q'_y\to 0)}{(2\pi L)^2}
\end{cases}
\end{equation}
%\AFnote{devo dire che non mi fa impazzire l'uso di $x,y$ per le due direzioni, perche' danno un senso molto coordinate cartesiane...}
%\AG{For the case of strong anisotropic systems, this approximation may depend on the specific orientation of the 2D system with respect to the periodic coordinate axis. If this is the case, the fully tensorial nature of $\bar{\bar{f}}_{\mathrm{lim}}$ should be taken into account.}
%\AG{We note that approximating $\bar{\bar{f}}_{\mathrm{lim}}$ to be diagonal may introduce a dependence on the reciprocal orientation between the 2D system and the coordinate system \DVnote{questo così non lo capisco, è necessario?}\AGnote{A me sembrava che questa nota fosse importante per Andrea}. However, this spurious dependence vanishes for isotropic systems.}
%\AFnote{Penso che sia importante capire il punto, ma non starei a stressare la cosa ulteriormente... secondo me questa frase si puo' togliere.}

Otherwise, we may neglect the anisotropy of the auxiliary function adding the following approximation:  $f^{xx}_{\mathrm{lim}}\approx f^{yy}_{\mathrm{lim}} \equiv f_{\mathrm{lim}}$, where
\begin{equation}\label{Eq_flim_noanis}
f_{\mathrm{lim}} = \frac{\scrc_{00}(\q'\to 0)}{(2\pi L)^2} \ .
\end{equation}
In Eq.~\eqref{Eq_flim_noanis}, the limit is performed along the in-plane $110$ cartesian direction, in order to partially average between the $x$ and $y$  directions.\\
%\AFnote{qui x and y sembrano di nuovo le direzioni dei generatori.}
The $\alpha$ and $\beta$ coefficients in Eq. \eqref{Eq_f00_fit} are obtained by interpolation, using the nearest neighbours of the $\q = 0$ point. 
%\AG{
We note there are only two independent nearest-neighbor conditions to be applied, due to the symmetry property $f_{00}(\q)=f_{00}(-\q)$ [which can be derived from Eq.~\eqref{usl2003} with the symmetry property $\chi^0_{\G\G'}(\q)=\chi^0_{-\G'-\G}(-\q)$].
%}
%\AG{We note that $f_{00}(\mathbf{q})=f_{00}(\mathbf{-q})$, as $\vcoul_0(\q)=\vcoul_0(-\q)$ and $\scrc_{00}(\q)=\scrc_00(-\q)$ due to the condition  $f_{00}(\mathbf{q})=f_{00}(\mathbf{-q})$ given by time-reversal symmetry, there are only three independent conditions to be applied.}
%\AFnote{qui penso che tu non ti riferisca alla simmetria di inversione spaziale, che non necessariamente e' presente... oppure pensi a inversione/time-reversal ? oppure semplicemente alla simmetria di W per q-> -q ?}
%\AGcancel{\AG{Finally}, it is important to note that $f_{00}(\q \to 0) $ is the most relevant element in the integration of the self-energy~\cite{Rasmussen_2016}.
%\AFnote{quest'ultima frase e' un po' appesa...}}

%\AFnote{qui $f_{00}(\q)$ e' ha valore tensoriale, che dipenden dalla direzione, vicino a 0?  dalla Eq.(21) direi di si... se cosi' secondo me non dobbiamo usare $f_{00}(\q = 0) $, ma $f_{00}(\q \to 0) $.}

%========================
\section{Results}
\label{results}
%========================
%
\begin{figure*}%[hbtp]
\includegraphics[width= \textwidth]{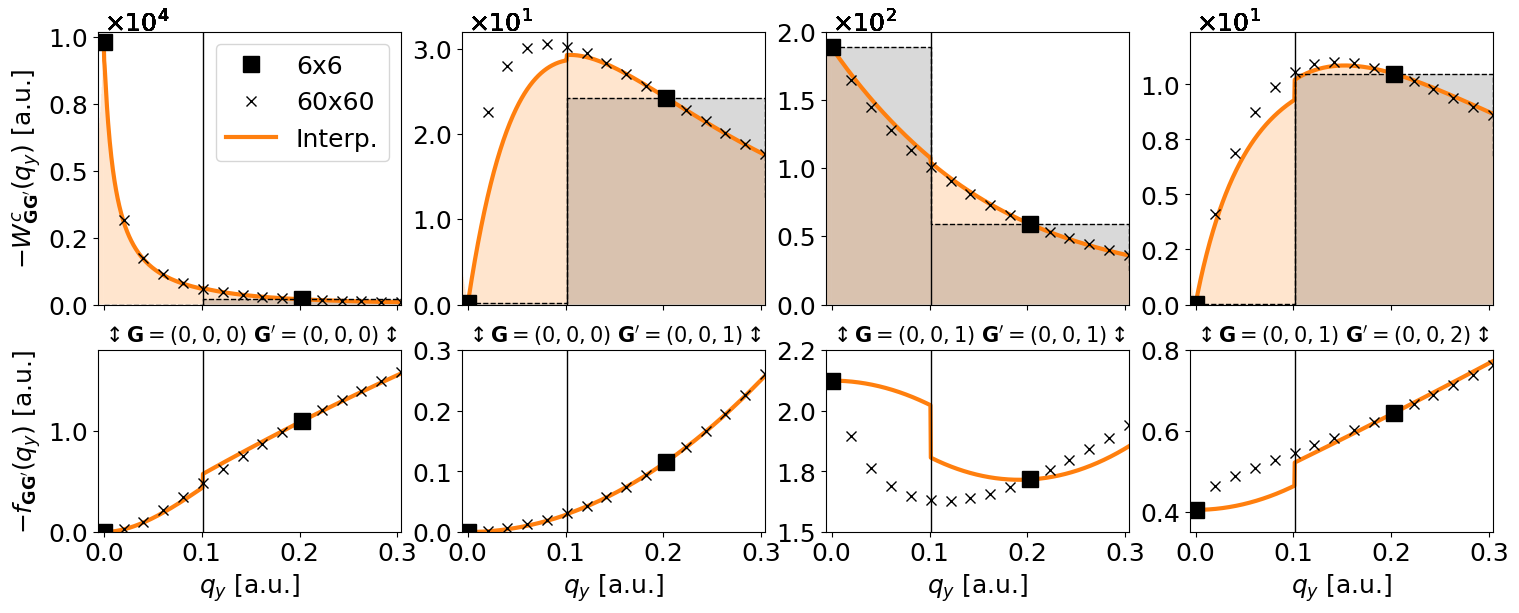}
\caption{Selected matrix elements of the correlation part of the screened potential (upper panels) and of the auxiliary function (lower panels) for the  MoS$_2$ monolayer.
Squares (crosses) are the numerical values obtained with a $6\times 6 \times 1$ ($60\times 60 \times 1$) Monkorst-Pack grid.
The interpolated functions between the points of the coarser mesh are plot with orange lines. The grey-shaded areas represent the values of the integral of $\scrc$ when a simple trapezoidal rule is applied to the coarser grid (See the text for details about the missing contribution for $\G=\G'=0$ and $q \approx 0$).
The orange-shaded area represent the values of the integral of $\scrc$ obtained with the W-av method.
Different domains of integration of the W-av method (which in 2D corresponds to the hexagons in the left panel of Fig.~\ref{fig_BZ}) are here separated with vertical black lines. We note the interpolation is discontinuous at the domain's border, as the interpolation procedure is applied at each domain separately.
%From left to right, we have: $\G = \G' = (0,0,0)$; $\G = (0,0,0)$, $\G' = (0,0,1)$; $\G = \G' = (0,0,1)$; $\G = (0,0,1)$, $\G' = (0,0,2)$ \DVnote{mettere nei panel}.
We remember that $z$ here is the non-periodic direction. 
The auxiliary functions are multiplied by a factor $10^2$ for clarity.
}
%\DVnote{mettere  W sopra e f sotto, inserire piu spazio fra plots fgg e W, mettere le componenti G,G' nei vari pannelli, eliminare MoS2 dall figura in mezzo: metterlo in alto a desrta fuori dalla figura o eliminarlo del tutto e inserirlo nella capcion. Legend: mettere punti square e corss con il relativo sampling. Ingrandire punti a q=0 per la griglia lasca. Dire nella caption cosa rappresenta la barra verticale (domains degli integrali?), se è così rimuoverle dali plots delle f}\AGnote{Done}
\label{W_fit}
\end{figure*}
We now present the results obtained with the interpolation scheme derived in Sec. \ref{Sec_methods}
for three prototype monolayer materials starting from the transition metal dicalcogenide MoS$_2$. Electronic properties of MoS$_2$ have been extensively studied in the literature, including several calculations using the $G_0W_0$ approach\cite{Qiu_2013,Huser_2013,Molina_2013,Qiu_2016}. It is a direct gap material with hexagonal structure having the gap localized at the $K/K'$ point in the BZ.
The valence band at $K/K'$ is split due to spin-orbit coupling~\cite{Zhu_2011}, but since we are interested in the convergence behavior with respect to the $\q$-point sampling, and for sake of simplicity, we have not included spin-orbit effects in the present calculations.
In addition, MoS$_2$ has been used to test two other convergence-accelerator schemes~\cite{Rasmussen_2016,Xia_2020}, which allows for a direct comparison with our approach.

In Fig.~\ref{W_fit}, we show some matrix elements of the correlation part of the screened potential $\scrc_{\G\G'}(\mathbf{q})$ 
%[see Eq.~\eqref{Eq_W}] 
and the auxiliary function $f_{\G\G'}$ [see Eq.~\eqref{Eq_f_num}] as a function of the momentum transfer $q$.
The interpolation functions (orange lines) 
are computed starting from the data on a coarse grid ($6 \times 6 \times 1$), and compared with the same quantities
%$f_{\G\G'}$ and $\scrc_{\G\G'}$ 
computed with a denser grid ($60 \times 60 \times 1 $), taken here as a benchmark.
The matrix element of $W$ contributing the most to the GW correction
is the $\G = \G' = 0$ term (Fig.~\ref{W_fit} left panels), being $\scrc_{\G\G'}$ at least two order of magnitude larger than the other elements.
%In fact, the calculation done with the dense grid shows that  $\scrc_{00}$ sharply approaches $\approx 10^4$ a.u. in the long-wavelength limit, while $\scrc_{\G\G'} \approx 10^{1}-10^{2}$ for the other matrix elements.
%This is the feature that is difficult to describe with standard discretization techniques and brings the convergence issues discussed in the previous sections.
As shown in Fig. \ref{W_fit}, for the $\G=\G'=0$ element there is a very good agreement between the results obtained interpolating the coarser grid (orange line) 
%with the interpolation functions 
and the values calculated using the denser grid.
%benchmark obtained using the dense grid, for $\G=\G'=0$.
For all the matrix elements considered, the auxiliary function $f_{\G\G'}$ is smoother than $\scrc_{\G\G'}$ which supports the choice of interpolating $f_{\G\G'}$ instead of $\scrc_{\G\G'}$.

Fig.~\ref{W_fit} shows no clear trend between the interpolation accuracy of $f_{\G\G'}$ and of $\scrc_{\G\G'}$, in particular in the region $q\approx 0$. %\DVnote{see comment}\AGnote{Answered}
%$\scrc_{\G\G'}$ is written in terms of $f_{\G\G'}$ through Eq.~\eqref{Eq_W_fit} via the bare Coulomb potential, given by Eq.~\eqref{Eq_cutoff_q}.
Since the different $\G$-components of the bare Coulomb potential, Eq.~\eqref{Eq_cutoff_q}, have different limits and slopes for $\bf q \to 0$, 
%and also approach the $\bf q \to 0$ limit with different slopes, 
the error associated with the interpolation of $f_{\G\G'}$ can be both enhanced or quenched when propagated to $\scrc_{\G\G'}$.
Despite this, we find a very good agreement between the interpolated values of $\scrc_{\G\G'}$ and the results obtained with the denser grid for all the considered matrix elements.

The gray shaded areas 
represent the integrals of $\scrc_{\G\G'}$ as obtained by applying the trapezoidal rule to the coarser grid together with the regularization of the Coulomb potential at $\G=\G'=\q=0$, given by Eq.~\eqref{Eq_v_av}.
For the sake of comparison, the same integrals, now obtained by using the interpolation, are highlighted in orange.
The trapezoidal rule, due to the regularization of the bare Coulomb potential, misses completely the integral contribution at $\G=\G'=0$ because of the vanishing value of $[\epsilon^{-1}_{00}(\q=0)-1]$, while $\bar{\vcoul}_0(\q=0)$ remains finite. 
%
%\DVcancel{In fact, as $\tilde{\epsilon}^{-1}_{\G\G'}(\q=0) = 0$, we find $\tilde{\epsilon}^{-1}_{\G\G'}(\q=0)\bar{\vcoul}_0(\q=0) = 0$.}
Therefore, averaging the whole $\scrc$, as we do in Eq.~\eqref{Eq_W_av}, instead of averaging $\vcoul$, Eq.~\eqref{Eq_v_av}, and multiplying by $[\epsilon^{-1}_{00}(\q)-1]$, is mandatory to have a contribution different from zero in this region.
%\AFnote{dobbiamo definire $\tilde{\epsilon}^{-1}_{\G\G'}(\q)$, che non e' piu' definito... c'e' una qualche notazione piu' standard ? usiamo questa ?}

We also note  that the trapezoidal rule misses the integral contributions of $\scrc_{\G\G'}$ for $\G = 0$ or $\G'=0$ (wings) in the long-wavelength limit ($q \to 0$), since
$\scrc_{\G\G'}(\q) \to 0$ as $q \to 0$.
%\AG{In fact, for those matrix elements, $\tilde{\epsilon}^{-1}_{\G\G'} \to 0$ and $\sqrt{\vcoul_{\G}}\sqrt{v_{\G'}} \to \infty$ in the long-wavelength limit (see Fig.~\ref{Fig_epsm1}), while $W_{\G\G'} = \sqrt{\vcoul_{\G}}\tilde{\epsilon}_{\G\G'}^{-1}\sqrt{\vcoul_{\G'}} \to $ constant.
%Thus, if the bare Coulomb interaction is averaged [e.g. via Eq.\eqref{Eq_reg_v_simple} or Eq.\eqref{Eq_v_av} for $\G=\G'$], then multiplied by the inverse dielectric function, the erroneous result $W_{\G\G'} = 0$ is obtained.}
Finally, when $\G,\G' \neq 0$, the trapezoidal rule overestimate the integral in the region $q \approx 0$.
The orange areas, obtained with the interpolation functions, give instead a good description of the  areas under the dense grid data.
This  justifies the accuracy of the interpolation method with fairly coarse grids, as detailed in the following.

\begin{figure}%[hbtp]
\includegraphics[width=0.5 \textwidth]{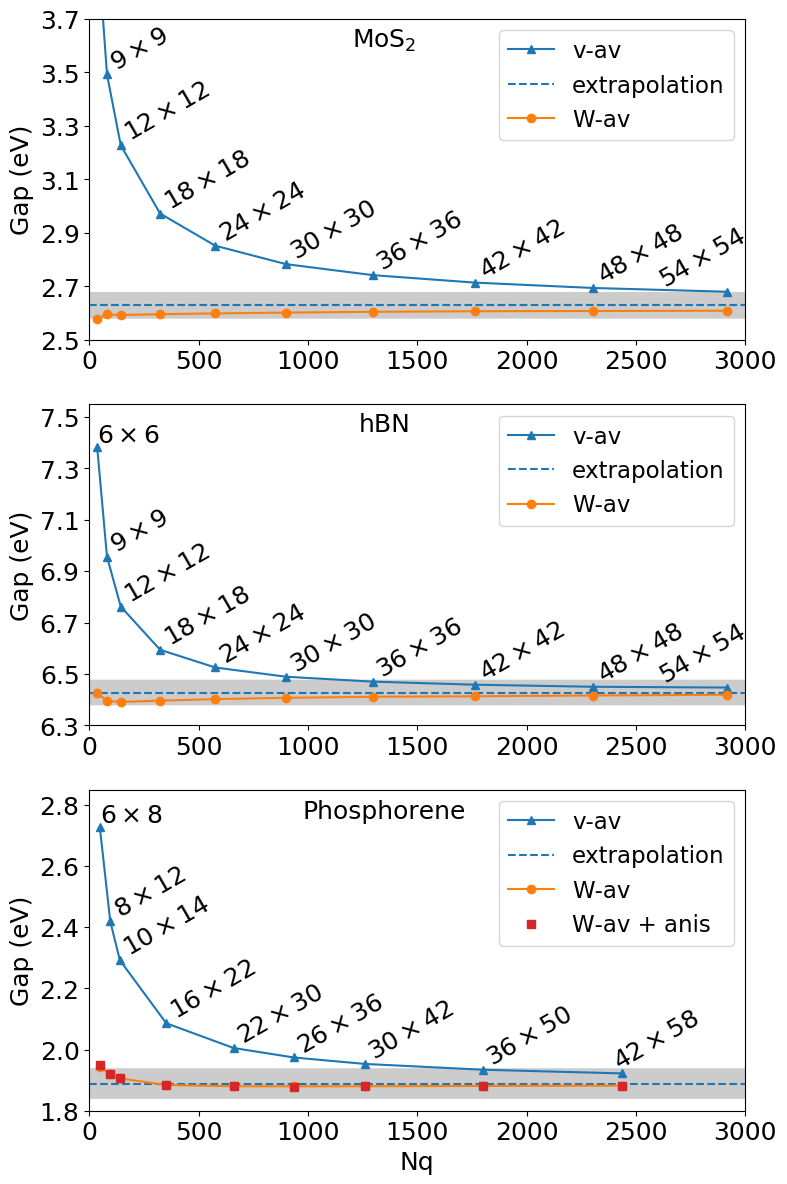}
\caption{Convergence of the quasiparticle band gap of MoS$_2$ (upper panel), hBN (middle panel) and phosphorene (lower panel) with respect to the number of sampling points of the BZ.
Blue lines indicate results obtained with standard integration methods.
The extrapolated values are indicated with an horizontal dashed line (see the text for more details about the extrapolation procedure).
Orange lines are the results obtained with the W-av method.
For the case of phosphorene, red squares indicate results obtained with the W-av method, in which the anisotropic behavior of the screened potential is included through Eq.~\eqref{Eq_flim_noanis}.
The grey shaded regions show the converge tolerance ($\pm 50$ meV) and are centered at the extrapolated values.
%\DVnote{Sostituire standard con trapezoidal rule. Sostituire Ph con Phosphorene}\AGnote{Done}
%
}
\label{Gap_conv}
\end{figure}

In Fig.~\ref{Gap_conv} we show the results for the QP band gap as a function of the {\bf q}-point sampling for 
%of three prototypical 2D semiconductors: 
MoS$_2$, hBN, and phosphorene.
%MoS$_2$ has been discussed above.
As for the case of MoS$_2$, also hBN~\cite{Ferreira_2019,Cudazzo_2016,Galvani_2016,Berseneva_2013,Ciraci_2009,Blase_1995,Wirtz_2006} and phosphorene monolayer~\cite{Li_2017,Qiu_2017,Yoon_2021} have been extensively studied using GW approach. 
%is a wide gap insulator, with an indirect gap at $K \to \Gamma$. 
%Due to the large value of the gap, it has weak screening properties, and is employed as spacer layer in artificially designed layered heterostructures[cite].
%
%The monolayer of phosphorene, also extensively studied\cite{Li_2017,Qiu_2017,Yoon_2021}, has a direct band gap located at the $\Gamma$ point in the BZ.
Moreover, due to its high anisotropy, the phosphorene monolayer is an ideal system to test the two proposed treatments of the $f_{\G\G'}$ anisotropy, given in Eqs.~\eqref{Eq_flim_noanis} and \eqref{Eq_flim_anis}.
%As in the case of MoS$_2$, the convergence properties of hBN and phosphorene with respect to the number of points in the $q$ sampling have already been studied [cite].

In Fig.~\ref{Gap_conv}, the convergence of the fundamental gap for the three materials as a function of the $\q$-sampling is shown using the proposed accelerated method (W-av) and 
the v-av method. In the latter case, only the $\q=\G=0$ term of the Coulomb interaction has been averaged, in order to regularize the Coulomb divergence.
We verified that the use of the v-av method to treat the $\q \neq 0$ an $\G \neq 0$ terms of the Coulomb interaction does not  significantly affect the results of the fundamental gap for the  considered systems.
The v-av method shows a very slow convergence with respect to $N_k$, as expected, and the gaps in the limit of an infinitely dense grids have been obtained by 
%\PDAcancel{extrapolating} 
using an $1/N_k$ 
%\PDAchange{fit}
extrapolation. 
For all the three cases we note that the gap is overestimated when unconverged grids are used, mainly due to the lack of the long wavelength contributions of the correlation parts of the screened potential, as explained in the previous section (see Fig.~\ref{W_fit}).
Using the v-av method, to obtain a gap value within less than $\pm 50$ meV with respect to the extrapolated value, 
%with the standard method, 
$\mathbf{k}$-grids of $54\times 54 \times 1$, $36\times 36\times 1$, and $36 \times 50 \times 1$ are required for MoS$_2$, hBN, and phosphorene, respectively.

With the proposed W-av method, the convergence of the gaps is greatly accelerated, and we obtain converged results already using  $6\times 6 \times 1$, $6\times 6\times 1$ and $8 \times 12 \times 1$ grids for MoS$_2$, hBN and phosphorene,  respectively, comparable with those required to obtain converged DFT results.
These grids are respectively 80, 40 and 20 times smaller than the ones required to have similar accuracy without acceleration.
%\DVnote{see comments}\AGnote{Answered}
%As DFT results are commonly used as the starting point of GW calculations, we obtained the highest possible convergence acceleration.
Converged results using similar size of {$\k$}-grids were also %These convergence acceleration levels have been already 
obtained with alternative accelerator schemes~\cite{daJornada_2017,Xia_2020}.
However, within the present method, differently from the other proposed strategies~\cite{daJornada_2017,Xia_2020}, no additional sub-sampling points are required to be computed in the region $q \approx 0$. 
%, as opposite to the methods described in Refs. [cite].

The orange dots in Fig.~\ref{Gap_conv} are obtained with a parametrization of the head of the auxiliary function given by Eq.~\eqref{Eq_flim_noanis}, which accounts for the anisotropy of the system by simply interpolating along the direction (110).
Nevertheless,
for phosphorene, that is highly anisotropic, we have also taken explicitly into account the anisotropy of $W^c$, using a parametrization of the auxiliary function given by Eq.~\eqref{Eq_flim_anis} (red dots). Although the long-wavelength limits of $\scrc_{00}$ are different, the average of the correlation part of the potential, see Eq.~\eqref{Eq_W_av}, is very similar in the two schemes and 
the resulting quasiparticle corrections do not show substantial differences.
%results 
%obtained considering or not the explicit anisotropic effects show similar computed quasiparticle gaps.
%In fact, although the long-wavelength limits are different, the average of the correlation part of the potential [given in Eq.~\eqref{Eq_W_av}] is very similar in the two schemes.
%\PDAnote{La seguente frase contiene una ripetizione abbastanza ravvicinata, la giro come segue, che ne dici?} 
Despite the present results for phosphorene show that the explicit anisotropic treatment does not affect the value of the computed band gap, 
%\PDAchange{the explicit treatment of the anisotropy is}
this does not exclude the fact that it can be potentially relevant for other systems and deserves further investigation.
%\AFcancel{and it has been implemented in the code.}
%
\begin{figure}%[hbtp]
\includegraphics[width=0.5 \textwidth]{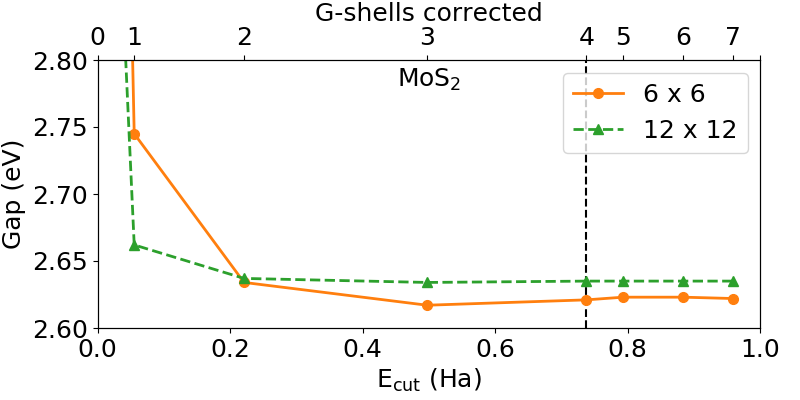}
\caption{Convergence of the quasiparticle band gap of MoS$_2$ obtained with the W-av method with respect to the cutoff energy of the correction. In the upper x-axis it is shown the numbers of G shells corrected.
Orange (green) lines indicate results obtained with the $6\times 6 \times 1$ ($12\times 12\times 1$) Monkorst-Pack grid.
The vertical dashed line indicates the $E^{av}_{cut}$ used in Fig.~\ref{Gap_conv}.
%\DVcancel{In the calculations reported in Fig.~\ref{Gap_conv}, we used a cutoff of $E^{av}_{cut} = min|\mathbf{G}_{xy}|^2/2$, here indicated with a vertical dashed line.}
}
\label{conv_ng}
\end{figure}

Next, we turn the attention on the role of the  
%Lets now consider in more detail the convergence of the proposed method with respect to the 
number of matrix elements of $\scrc_{\G\G'}(\mathbf{q})$ averaged through Eq.~\eqref{Eq_W_av}, identified by the parameter $E_{cut}^{av}$.
In Fig.~\ref{conv_ng}, we plot the convergence of the band gap of MoS$_2$ with respect to  $E_{cut}^{av}$, or, alternatively, with respect to the number of $\G$ shells for which the averaging procedure is employed.
In the plot, points at $E_{cut}^{av} = 0$ refer to gaps obtained with the v-av method, i.e. blue triangles shown in the top panel of Fig.~\ref{Gap_conv}.
%in which no correction is applied to $\scrc_{\G\G'}$.
The W averaging of the first element gives the largest contribution to the  convergence acceleration, closing the gap to $1.28$ and $0.60$ eV for the $6\times 6 \times 1$ and $12 \times 12 \times 1$ grids, respectively.
The W averaging of the first $\G_{\perp}$ matrix elements is also important to obtain converged results with coarser grids.
In particular, the coarser the grid, the more important is the averaging of $\scrc_{\G_{\perp}\G'_{\perp}}$, as shown by the comparison of the $6 \times 6 \times 1$ with the $12 \times 12 \times 1$ grids.
%\AFnote{ho cancellato il commento per errore, sorry... Riguardo le griglie, mi stavo chiedendo se il punto non sia che con la 12x12 possiamo fare le cose un po' peggio, perche' tanto siamo intrinsecamente piu' accurati...}
Still, in both cases, the convergence is reached for a small number of G shells, which translates into a nearly negligible added computational cost %\AGcancel{with respect to standard discretization techniques} 
required to perform the averaging of the screened potential. 
According to our results, $E_{cut}^{av} \approx 1$-$2$ Ry %\AFnote{$\G_\parallel$ ?} 
%\AFcancel{seems to be} 
is a reasonable choice for all the systems considered.

%and was indeed the value employed in the calculations shown in Fig.~\ref{Gap_conv} \DVnote{see comment}. \AFnote{anche secondo me non male spiegare meglio il razionale}

%========================
\section{Computational details}
\label{compdet}
%========================
%
DFT calculations were performed using a plane wave basis set as implemented in the Quantum ESPRESSO package~\citep{QE_2020}, and
using the Perdew-Burke-Ernzerhof (PBE) exchange-correlation functional~\citep{PBE}.
%The following lattice constants have been obtained after the structure relaxation: $a=3.159$ Angstrom for MoS$_2$, $a = 2.509$ Angstrom for hBN, and $a=?$ Angstrom for phosphorene.
We have considered supercells with an interlayer distance $L = 10$ \AA \ for MoS$_2$ and $L = 15$ \AA \ for hBN and phosphorene, which are enough to obtain converged results with respect to the cell size, in agreement with Ref.~\onlinecite{Rasmussen_2016}.
%We numerically verified the convergence of the results with respect to $L$.
%Our findings are in agreement with the literature ~\citep{Rasmussen_2016}.
%The Kohn-Sham equations have been solved in reciprocal space with a plane-wave cutoff of $60$ Ry.
The kinetic energy cutoff for
the wavefunctions was set to $60$ Ry and we adopted norm-conserving pseudopotentials to model the electron–ion interaction.
%Then, single particle energies and wavefunction have been used as input for $G_0W_0$ calculations.\\

$G_0W_0$ calculations were performed with the Yambo package~\citep{yambo_2009,yambo_2019}. We used a cutoff of $5$ Ry for the size of the dielectric matrix, including up to $400$ states in the sum-over-state of the response function.
The same number of states has been employed in the calculation of the correlation part of the self energy.
%\AFnote{quante bande nella sum-over-states della Sigma?} 
To accelerate the convergence with respect to the number of empty states we have used the algorithm described in Refs.~\onlinecite{Bruneval_2008,yambo_2019}.
We have numerically verified the convergence of the QP gap up to 50 meV with respect to the number of bands included in the calculation of both the Green and polarization functions.
%We have used G-type terminators~\citep{Bruneval_2008,yambo_2019} to accelerate the convergence with respect to the number of unoccupied states.
%response functions have been obtained with a cutoff of $5$ Ry in reciprocal space.
Despite the cutoff used to represent the dielectric matrix is not enough to provide highly converged QP properties, it is sufficient to provide accurate convergence trends with respect to the $\mathbf{q}$-sampling.
%In fact, high energy matrix elements of the response functions converge very rapidly with respect to the q mesh.

%Sums over states have been \AG{performed including} the first $400$ states.
%We have used G-type terminators~\citep{Bruneval_2008,yambo_2019} to accelerate the convergence with respect to the number of unoccupied states.
%\DVnote{to be checked}\AGnote{Checked, see the additional comment}.

%========================
\section{Conclusions}
\label{concl}
%========================
%
%We succeed in the contruction of a method to accelerate the convergence of GW results with respect to the number of k point sampling of the BZ.
Accurate results for the calculation of quasiparticle energies in the GW approximation for 2D semiconductors can be obtained only by using very large $\mathbf{k}$-point grids, making  calculations computationally very demanding.  
We have provided here a technique based on a stochastic averaging and interpolation of the screened potential to accelerate the convergence of the self energy with respect to the $\mathbf{q}$-point sampling.
%momentum transfer integrals in the calculation of the self-energies.
We have tested the proposed scheme for the calculation of the QP gap of three prototypical monolayer semiconductors: MoS$_2$, hBN, and phosphorene.
We find that grids such as $6\times 6 \times 1$, $6\times 6\times 1$ and $8 \times 12 \times 1$ are enough to obtain converged results for the fundamental gap up to $50$ meV for MoS$_2$, hBN, and phosphorene, respectively. These grids are 80,40 and 20 times smaller than those required to achieve a similar accuracy when averaging only the bare coulomb potential (v-av method).
Taking the $\mathbf{k}$ and $\mathbf{q}$ grids to be identical, $G_0W_0$ typically scales\cite{martin_reining_ceperley_2016} as $N_k^2$. When this is the case, with the proposed method the computational cost of a $G_0W_0$ calculation is reduced by at least two orders of magnitude, without loss of accuracy. 
%\DVnote{see comment}\AGnote{Answered}
%\DVnote{see comment again}\AFnote{ok ?}
%
%With standard integration techniques, the $54\times 54 \times 1$, $36\times 36\times 1$ and $36 \times 50 \times 1$ are required instead to converge the gap for MoS$_2$, hBN and phosphorene respectively.
%\PDAnote{forse anche qui si possono aggiungere i fattori moltiplicativi che mostrano di quanto scala il calcolo rispetto al metodo standard (come nell'abstract), da cui poi deriva il fattore 100 che segue}.
The proposed W-av method is able to describe the anisotropy of the screened potential at different levels of approximations, and differently from other methods recently proposed does not rely on any sub-sampling of the BZ.
%by changing the model function which parametrize the auxiliary function $f_{00}$.

%This work can be easily extended to systems with different dimensionalities, from 1D to 3D, by changing the expression of the Coulomb potential to be appropriately truncated.
The possibility to extend the present methodology to metals and systems with different dimensionalities (1D or 3D) is envisaged and will be explored in a future research.

%========================
\section*{Acknowledgments}
%========================
%
We acknowledge stimulating discussions with Dario A. Leon, Miki Bonacci, Simone Vacondio and Matteo Zanfrognini. This work was partially supported by SUPER (Supercomputing Unified Platform – Emilia-Romagna) from Emilia-Romagna PORFESR 2014-2020 regional funds. We also thank MaX -- MAterials design at the eXascale -- a European Centre of Excellence, funded by the European Union program H2020-INFRAEDI-2018-1 (Grant No. 824143). Computational time on the Galileo machine at CINECA was provided by the Italian ISCRA program.

\appendix

%========================
\section{Head matrix element of the screened potential in terms of auxiliary function}
\label{Appx1}
%========================
%In this appendix, we derive Eq. \eqref{Eq_Dec_W}.
%\DVchange{In this appendix we derive the expression for the head of the screened potential following the work by Rasmussen et al.~\cite{Rasmussen_2016}}
{We will follow a derivation similar to the one adopted by Rasmussen et al.~\cite{Rasmussen_2016},} but without resorting to the long-wavelength limit.
The symmetrized static dielectric function $\eps$ is expressed in terms of the irreducible static response function $\irrchi$ as
\begin{equation}\label{usl2000}
\eps_{\G\G'}(\q) = \delta_{\G,\G'}-\sqrt{\vcoul_{\G}(\q)}\irrchi_{\G\G'}(\q)\sqrt{\vcoul_{\G'}(\q)} \ ,
\end{equation}
where $\vcoul_{\G}$ is the Coulomb interaction given by Eq. \eqref{Eq_cutoff_q}.
The inverse dielectric function can be expressed by considering $\eps$ as a block matrix in the $\G$, $\G'$ components with head ($\G=\G'=0$) ($H$), wings ($\G =0, \G' \neq 0$  and $\G \neq 0, \G' = 0$) ($\mathbf{w}$ and $\mathbf{v}$), and body ($\G \neq 0, \G' \neq 0$) ($\mathbf{B}$):
%\AFnote{visto che prendiamo tutto simmetrized, perche' dobbiamo usare $v$ e $w$ come simboli diversi ? e' necessario ? A me la Xo statica sembra proprio hermitiana anche a q finito. Mi pare cmq che questo non abbia conseguenze.}
%
\begin{equation}\label{Eqepsblock}
\eps = \left(\mqty{H & \mathbf{w}^{\mathrm{T}} \\ \mathbf{v} & \mathbf{B}}\right) \ .
\end{equation}
The inverse matrix reads: 
%\DVcancel{is then given by inverting the block matrix}
\begin{widetext}
\begin{equation}
\inveps = \left(\mqty{(H-\mathbf{w}^{\mathrm{T}}\mathbf{B}^{-1}\mathbf{v})^{-1} &
-(H-\mathbf{w}^{\mathrm{T}}\mathbf{B}^{-1}\mathbf{v})^{-1}\mathbf{w}^{\mathrm{T}}\mathbf{B}^{-1}\\
-\mathbf{B}^{-1}\mathbf{v}(H-\mathbf{w}^{\mathrm{T}}\mathbf{B}^{-1}\mathbf{v})^{-1} & \mathbf{B}^{-1}+\mathbf{B}^{-1}\mathbf{v}(H-\mathbf{w}^{\mathrm{T}}\mathbf{B}^{-1}\mathbf{v})^{-1}\mathbf{w}^{\mathrm{T}}\mathbf{B}^{-1}}\right) \ .
\end{equation}
We now focus on the head of the inverse dielectric function
\begin{equation}\label{usl2001}
\inveps_{00}(\q) = \left(\eps_{00}(\q)-\sum\limits_{\G\G' \neq 0}\eps_{0\G}(\q)\mathbf{B}_{\G\G'}^{-1}(\q)\eps_{\G'0}(\q)\right)^{-1} \ .
\end{equation}
By substituting Eq. \eqref{usl2000} into Eq. \eqref{usl2001}, we find:
\begin{equation}\label{usl2002}
\inveps_{00}(\q) = \left[1-\vcoul_0(\q)\left(\irrchi_{00}(\q)-\sum\limits_{\G\G' \neq 0}\irrchi_{0\G}(\q)\sqrt{\vcoul_{\G}(\q)}\mathbf{B}_{\G\G'}^{-1}(\q)\sqrt{\vcoul_{\G'}(\q)}\irrchi_{\G'0}(\q)\right)\right]^{-1} \ .
\end{equation}
We define the $f$ function as
\begin{equation}\label{usl2003}
f(\q) \equiv 
\irrchi_{00}(\q)-\sum\limits_{\G\G'}\irrchi_{0\G}(\q)\sqrt{\vcoul_{G}(\q)}\mathbf{B}^{-1}_{\G\G'}(\q)\sqrt{\vcoul_{G'}(\q)}\irrchi_{\G'0}(\q) \ ,
\end{equation}
\end{widetext}
%If local fields are neglected, $f(\q) = \irrchi_{00}(\q)$.
%This physical consideration may suggest the functional form to be used for the interpolation of $f$.
%For example, $f(\q \to 0) \to 0$ as $|\q|^2$, as $\irrchi_{00}(\q) \to 0$ as $|\q|^2$.\\
By substituting Eq. \eqref{usl2003} into Eq. \eqref{usl2002}, we have
\begin{equation}\label{usl3002}
\inveps_{00}(\q) = \frac{1}{1-\vcoul_0(\q)f(\q)} \ .
\end{equation}
%Thus, the correlation part of the inverse dielectric function is
%\begin{equation}
%\invepsc_{00}(\q) = \inveps_{00}(\q)-1 = \frac{\vcoul_0(\q)f(\q)}{1-\vcoul_0(\q)f(\q)} \ ,
%\end{equation}
By substituting Eq. \eqref{usl3002}\ into Eq.~\eqref{Eq_W} with $\G = \G' = \mathbf{0}$, the correlation part of the static screened potential is
\begin{equation}\label{usl3003}
\scrc_{00}(\q) = [\inveps_{00}(\q)-1]\vcoul_0(\q) = \frac{\vcoul_0(\q)f(\q)\vcoul_0(\q)}{1-\vcoul_0(\q)f(\q)} \ .
\end{equation}
An equation with the same structure as Eq.~\eqref{usl3003} can be derived for each diagonal element of the screening ($\G=\G' \neq 0$) by rearranging the lattice vectors so that the first line and column correspond to a $\G \neq 0$.

\bibliography{bibliography}

\begin{thebibliography}{64}%
\makeatletter
\providecommand \@ifxundefined [1]{%
 \@ifx{#1\undefined}
}%
\providecommand \@ifnum [1]{%
 \ifnum #1\expandafter \@firstoftwo
 \else \expandafter \@secondoftwo
 \fi
}%
\providecommand \@ifx [1]{%
 \ifx #1\expandafter \@firstoftwo
 \else \expandafter \@secondoftwo
 \fi
}%
\providecommand \natexlab [1]{#1}%
\providecommand \enquote  [1]{``#1''}%
\providecommand \bibnamefont  [1]{#1}%
\providecommand \bibfnamefont [1]{#1}%
\providecommand \citenamefont [1]{#1}%
\providecommand \href@noop [0]{\@secondoftwo}%
\providecommand \href [0]{\begingroup \@sanitize@url \@href}%
\providecommand \@href[1]{\@@startlink{#1}\@@href}%
\providecommand \@@href[1]{\endgroup#1\@@endlink}%
\providecommand \@sanitize@url [0]{\catcode `\\12\catcode `\$12\catcode
  `\&12\catcode `\#12\catcode `\^12\catcode `\_12\catcode `\%12\relax}%
\providecommand \@@startlink[1]{}%
\providecommand \@@endlink[0]{}%
\providecommand \url  [0]{\begingroup\@sanitize@url \@url }%
\providecommand \@url [1]{\endgroup\@href {#1}{\urlprefix }}%
\providecommand \urlprefix  [0]{URL }%
\providecommand \Eprint [0]{\href }%
\providecommand \doibase [0]{http://dx.doi.org/}%
\providecommand \selectlanguage [0]{\@gobble}%
\providecommand \bibinfo  [0]{\@secondoftwo}%
\providecommand \bibfield  [0]{\@secondoftwo}%
\providecommand \translation [1]{[#1]}%
\providecommand \BibitemOpen [0]{}%
\providecommand \bibitemStop [0]{}%
\providecommand \bibitemNoStop [0]{.\EOS\space}%
\providecommand \EOS [0]{\spacefactor3000\relax}%
\providecommand \BibitemShut  [1]{\csname bibitem#1\endcsname}%
\let\auto@bib@innerbib\@empty
%</preamble>
\bibitem [{\citenamefont {Hedin}(1965)}]{Hedin_1965}%
  \BibitemOpen
  \bibfield  {author} {\bibinfo {author} {\bibfnamefont {L.}~\bibnamefont
  {Hedin}},\ }\href {\doibase 10.1103/PhysRev.139.A796} {\bibfield  {journal}
  {\bibinfo  {journal} {Phys. Rev.}\ }\textbf {\bibinfo {volume} {139}},\
  \bibinfo {pages} {A796} (\bibinfo {year} {1965})}\BibitemShut {NoStop}%
\bibitem [{\citenamefont {Strinati}\ \emph {et~al.}(1982)\citenamefont
  {Strinati}, \citenamefont {Mattausch},\ and\ \citenamefont
  {Hanke}}]{Strinati_1982}%
  \BibitemOpen
  \bibfield  {author} {\bibinfo {author} {\bibfnamefont {G.}~\bibnamefont
  {Strinati}}, \bibinfo {author} {\bibfnamefont {H.~J.}\ \bibnamefont
  {Mattausch}}, \ and\ \bibinfo {author} {\bibfnamefont {W.}~\bibnamefont
  {Hanke}},\ }\href {\doibase 10.1103/PhysRevB.25.2867} {\bibfield  {journal}
  {\bibinfo  {journal} {Phys. Rev. B}\ }\textbf {\bibinfo {volume} {25}},\
  \bibinfo {pages} {2867} (\bibinfo {year} {1982})}\BibitemShut {NoStop}%
\bibitem [{\citenamefont {Hybertsen}\ and\ \citenamefont
  {Louie}(1986)}]{Hybertsen_1986}%
  \BibitemOpen
  \bibfield  {author} {\bibinfo {author} {\bibfnamefont {M.~S.}\ \bibnamefont
  {Hybertsen}}\ and\ \bibinfo {author} {\bibfnamefont {S.~G.}\ \bibnamefont
  {Louie}},\ }\href {\doibase 10.1103/PhysRevB.34.5390} {\bibfield  {journal}
  {\bibinfo  {journal} {Phys. Rev. B}\ }\textbf {\bibinfo {volume} {34}},\
  \bibinfo {pages} {5390} (\bibinfo {year} {1986})}\BibitemShut {NoStop}%
\bibitem [{\citenamefont {Godby}\ \emph {et~al.}(1988)\citenamefont {Godby},
  \citenamefont {Schl\"uter},\ and\ \citenamefont {Sham}}]{Godby_1988}%
  \BibitemOpen
  \bibfield  {author} {\bibinfo {author} {\bibfnamefont {R.~W.}\ \bibnamefont
  {Godby}}, \bibinfo {author} {\bibfnamefont {M.}~\bibnamefont {Schl\"uter}}, \
  and\ \bibinfo {author} {\bibfnamefont {L.~J.}\ \bibnamefont {Sham}},\ }\href
  {\doibase 10.1103/PhysRevB.37.10159} {\bibfield  {journal} {\bibinfo
  {journal} {Phys. Rev. B}\ }\textbf {\bibinfo {volume} {37}},\ \bibinfo
  {pages} {10159} (\bibinfo {year} {1988})}\BibitemShut {NoStop}%
\bibitem [{\citenamefont {Reining}(2018)}]{reining2018gw}%
  \BibitemOpen
  \bibfield  {author} {\bibinfo {author} {\bibfnamefont {L.}~\bibnamefont
  {Reining}},\ }\href@noop {} {\bibfield  {journal} {\bibinfo  {journal} {Wiley
  Interdisciplinary Reviews: Computational Molecular Science}\ }\textbf
  {\bibinfo {volume} {8}},\ \bibinfo {pages} {e1344} (\bibinfo {year}
  {2018})}\BibitemShut {NoStop}%
\bibitem [{\citenamefont {Golze}\ \emph {et~al.}(2019)\citenamefont {Golze},
  \citenamefont {Dvorak},\ and\ \citenamefont {Rinke}}]{Golze_2019}%
  \BibitemOpen
  \bibfield  {author} {\bibinfo {author} {\bibfnamefont {D.}~\bibnamefont
  {Golze}}, \bibinfo {author} {\bibfnamefont {M.}~\bibnamefont {Dvorak}}, \
  and\ \bibinfo {author} {\bibfnamefont {P.}~\bibnamefont {Rinke}},\ }\href
  {\doibase 10.3389/fchem.2019.00377} {\bibfield  {journal} {\bibinfo
  {journal} {Frontiers in Chem.}\ }\textbf {\bibinfo {volume} {7}},\ \bibinfo
  {pages} {377} (\bibinfo {year} {2019})}\BibitemShut {NoStop}%
\bibitem [{\citenamefont {Marzari}\ \emph {et~al.}(2021)\citenamefont
  {Marzari}, \citenamefont {Ferretti},\ and\ \citenamefont
  {Wolverton}}]{Marzari2021NatMaterials}%
  \BibitemOpen
  \bibfield  {author} {\bibinfo {author} {\bibfnamefont {N.}~\bibnamefont
  {Marzari}}, \bibinfo {author} {\bibfnamefont {A.}~\bibnamefont {Ferretti}}, \
  and\ \bibinfo {author} {\bibfnamefont {C.}~\bibnamefont {Wolverton}},\ }\href
  {\doibase 10.1038/s41563-021-01013-3} {\bibfield  {journal} {\bibinfo
  {journal} {Nature Materials}\ }\textbf {\bibinfo {volume} {20}},\ \bibinfo
  {pages} {736} (\bibinfo {year} {2021})}\BibitemShut {NoStop}%
\bibitem [{\citenamefont {Damascelli}\ \emph {et~al.}(2003)\citenamefont
  {Damascelli}, \citenamefont {Hussain},\ and\ \citenamefont
  {Shen}}]{Damascelli2003RMP}%
  \BibitemOpen
  \bibfield  {author} {\bibinfo {author} {\bibfnamefont {A.}~\bibnamefont
  {Damascelli}}, \bibinfo {author} {\bibfnamefont {Z.}~\bibnamefont {Hussain}},
  \ and\ \bibinfo {author} {\bibfnamefont {Z.-X.}\ \bibnamefont {Shen}},\
  }\href {\doibase https://doi.org/10.1103/RevModPhys.75.473} {\bibfield
  {journal} {\bibinfo  {journal} {Rev. Mod. Phys.}\ }\textbf {\bibinfo {volume}
  {75}},\ \bibinfo {pages} {473} (\bibinfo {year} {2003})}\BibitemShut
  {NoStop}%
\bibitem [{\citenamefont {Guzzo}\ \emph {et~al.}(2011)\citenamefont {Guzzo},
  \citenamefont {Lani}, \citenamefont {Sottile}, \citenamefont {Romaniello},
  \citenamefont {Gatti}, \citenamefont {Kas}, \citenamefont {Rehr},
  \citenamefont {Silly}, \citenamefont {Sirotti},\ and\ \citenamefont
  {Reining}}]{Guzzo2011PRL}%
  \BibitemOpen
  \bibfield  {author} {\bibinfo {author} {\bibfnamefont {M.}~\bibnamefont
  {Guzzo}}, \bibinfo {author} {\bibfnamefont {G.}~\bibnamefont {Lani}},
  \bibinfo {author} {\bibfnamefont {F.}~\bibnamefont {Sottile}}, \bibinfo
  {author} {\bibfnamefont {P.}~\bibnamefont {Romaniello}}, \bibinfo {author}
  {\bibfnamefont {M.}~\bibnamefont {Gatti}}, \bibinfo {author} {\bibfnamefont
  {J.~J.}\ \bibnamefont {Kas}}, \bibinfo {author} {\bibfnamefont {J.~J.}\
  \bibnamefont {Rehr}}, \bibinfo {author} {\bibfnamefont {M.~G.}\ \bibnamefont
  {Silly}}, \bibinfo {author} {\bibfnamefont {F.}~\bibnamefont {Sirotti}}, \
  and\ \bibinfo {author} {\bibfnamefont {L.}~\bibnamefont {Reining}},\ }\href
  {\doibase https://doi.org/10.1103/PhysRevLett.107.166401} {\bibfield
  {journal} {\bibinfo  {journal} {Phys. Rev. Lett.}\ }\textbf {\bibinfo
  {volume} {107}},\ \bibinfo {pages} {166401} (\bibinfo {year}
  {2011})}\BibitemShut {NoStop}%
\bibitem [{\citenamefont {Caruso}\ \emph {et~al.}(2018)\citenamefont {Caruso},
  \citenamefont {Verdi}, \citenamefont {Ponc\'e},\ and\ \citenamefont
  {Giustino}}]{Caruso2018PRB}%
  \BibitemOpen
  \bibfield  {author} {\bibinfo {author} {\bibfnamefont {F.}~\bibnamefont
  {Caruso}}, \bibinfo {author} {\bibfnamefont {C.}~\bibnamefont {Verdi}},
  \bibinfo {author} {\bibfnamefont {S.}~\bibnamefont {Ponc\'e}}, \ and\
  \bibinfo {author} {\bibfnamefont {F.}~\bibnamefont {Giustino}},\ }\href
  {\doibase https://doi.org/10.1103/PhysRevB.97.165113} {\bibfield  {journal}
  {\bibinfo  {journal} {Phys. Rev. B}\ }\textbf {\bibinfo {volume} {97}},\
  \bibinfo {pages} {165113} (\bibinfo {year} {2018})}\BibitemShut {NoStop}%
\bibitem [{\citenamefont {Marini}\ \emph {et~al.}(2002)\citenamefont {Marini},
  \citenamefont {Del~Sole}, \citenamefont {Rubio},\ and\ \citenamefont
  {Onida}}]{Marini2002PRB}%
  \BibitemOpen
  \bibfield  {author} {\bibinfo {author} {\bibfnamefont {A.}~\bibnamefont
  {Marini}}, \bibinfo {author} {\bibfnamefont {R.}~\bibnamefont {Del~Sole}},
  \bibinfo {author} {\bibfnamefont {A.}~\bibnamefont {Rubio}}, \ and\ \bibinfo
  {author} {\bibfnamefont {G.}~\bibnamefont {Onida}},\ }\href {\doibase
  https://doi.org/10.1103/PhysRevB.66.161104} {\bibfield  {journal} {\bibinfo
  {journal} {Phys. Rev. B}\ }\textbf {\bibinfo {volume} {66}},\ \bibinfo
  {pages} {161104} (\bibinfo {year} {2002})}\BibitemShut {NoStop}%
\bibitem [{\citenamefont {Bechstedt}\ \emph {et~al.}(1994)\citenamefont
  {Bechstedt}, \citenamefont {Fiedler}, \citenamefont {Kress},\ and\
  \citenamefont {Del~Sole}}]{Bechstedt1994PRB}%
  \BibitemOpen
  \bibfield  {author} {\bibinfo {author} {\bibfnamefont {F.}~\bibnamefont
  {Bechstedt}}, \bibinfo {author} {\bibfnamefont {M.}~\bibnamefont {Fiedler}},
  \bibinfo {author} {\bibfnamefont {C.}~\bibnamefont {Kress}}, \ and\ \bibinfo
  {author} {\bibfnamefont {R.}~\bibnamefont {Del~Sole}},\ }\href {\doibase
  https://doi.org/10.1103/PhysRevB.49.7357} {\bibfield  {journal} {\bibinfo
  {journal} {Phys. Rev. B}\ }\textbf {\bibinfo {volume} {49}},\ \bibinfo
  {pages} {7357} (\bibinfo {year} {1994})}\BibitemShut {NoStop}%
\bibitem [{\citenamefont {Zhou}\ \emph {et~al.}(2020)\citenamefont {Zhou},
  \citenamefont {Reining}, \citenamefont {Nicolaou}, \citenamefont {Bendounan},
  \citenamefont {Ruotsalainen}, \citenamefont {Vanzini}, \citenamefont {Kas},
  \citenamefont {Rehr}, \citenamefont {Muntwiler}, \citenamefont {Strocov},
  \citenamefont {Sirotti},\ and\ \citenamefont {Gatti}}]{Gatti2020PNAS}%
  \BibitemOpen
  \bibfield  {author} {\bibinfo {author} {\bibfnamefont {J.~S.}\ \bibnamefont
  {Zhou}}, \bibinfo {author} {\bibfnamefont {L.}~\bibnamefont {Reining}},
  \bibinfo {author} {\bibfnamefont {A.}~\bibnamefont {Nicolaou}}, \bibinfo
  {author} {\bibfnamefont {A.}~\bibnamefont {Bendounan}}, \bibinfo {author}
  {\bibfnamefont {K.}~\bibnamefont {Ruotsalainen}}, \bibinfo {author}
  {\bibfnamefont {M.}~\bibnamefont {Vanzini}}, \bibinfo {author} {\bibfnamefont
  {J.~J.}\ \bibnamefont {Kas}}, \bibinfo {author} {\bibfnamefont {J.~J.}\
  \bibnamefont {Rehr}}, \bibinfo {author} {\bibfnamefont {M.}~\bibnamefont
  {Muntwiler}}, \bibinfo {author} {\bibfnamefont {V.~N.}\ \bibnamefont
  {Strocov}}, \bibinfo {author} {\bibfnamefont {F.}~\bibnamefont {Sirotti}}, \
  and\ \bibinfo {author} {\bibfnamefont {M.}~\bibnamefont {Gatti}},\ }\href
  {\doibase https://doi.org/10.1073/pnas.2012625117} {\bibfield  {journal}
  {\bibinfo  {journal} {Proc. Nat. Ac. Sci.}\ }\textbf {\bibinfo {volume}
  {117}},\ \bibinfo {pages} {28596} (\bibinfo {year} {2020})}\BibitemShut
  {NoStop}%
\bibitem [{\citenamefont {Geim}\ and\ \citenamefont
  {Novoselov}(2010)}]{geim2010rise}%
  \BibitemOpen
  \bibfield  {author} {\bibinfo {author} {\bibfnamefont {A.~K.}\ \bibnamefont
  {Geim}}\ and\ \bibinfo {author} {\bibfnamefont {K.~S.}\ \bibnamefont
  {Novoselov}},\ }in\ \href@noop {} {\emph {\bibinfo {booktitle} {Nanoscience
  and technology: a collection of reviews from nature journals}}}\ (\bibinfo
  {publisher} {World Scientific},\ \bibinfo {year} {2010})\ pp.\ \bibinfo
  {pages} {11--19}\BibitemShut {NoStop}%
\bibitem [{\citenamefont {Ferrari}\ \emph {et~al.}(2015)\citenamefont
  {Ferrari}, \citenamefont {Bonaccorso}, \citenamefont {Fal{'}ko},
  \citenamefont {Novoselov}, \citenamefont {Roche}, \citenamefont {Boggild},
  \citenamefont {Borini}, \citenamefont {Koppens}, \citenamefont {Palermo},
  \citenamefont {Pugno}, \citenamefont {Garrido}, \citenamefont {Sordan},
  \citenamefont {Bianco}, \citenamefont {Ballerini}, \citenamefont {Prato},
  \citenamefont {Lidorikis}, \citenamefont {Kivioja}, \citenamefont
  {Marinelli}, \citenamefont {Ryhänen}, \citenamefont {Morpurgo},
  \citenamefont {Coleman}, \citenamefont {Nicolosi}, \citenamefont {Colombo},
  \citenamefont {Fert}, \citenamefont {Garcia-Hernandez}, \citenamefont
  {Bachtold}, \citenamefont {Schneider}, \citenamefont {Guinea}, \citenamefont
  {Dekker}, \citenamefont {Barbone}, \citenamefont {Sun}, \citenamefont
  {Galiotis}, \citenamefont {Grigorenko}, \citenamefont {Konstantatos},
  \citenamefont {Kis}, \citenamefont {Katsnelson}, \citenamefont {Vandersypen},
  \citenamefont {Loiseau}, \citenamefont {Morandi}, \citenamefont {Neumaier},
  \citenamefont {Treossi}, \citenamefont {Pellegrini}, \citenamefont {Polini},
  \citenamefont {Tredicucci}, \citenamefont {Williams}, \citenamefont
  {Hee~Hong}, \citenamefont {Ahn}, \citenamefont {Min~Kim}, \citenamefont
  {Zirath}, \citenamefont {van Wees}, \citenamefont {van~der Zant},
  \citenamefont {Occhipinti}, \citenamefont {Di~Matteo}, \citenamefont
  {Kinloch}, \citenamefont {Seyller}, \citenamefont {Quesnel}, \citenamefont
  {Feng}, \citenamefont {Teo}, \citenamefont {Rupesinghe}, \citenamefont
  {Hakonen}, \citenamefont {Neil}, \citenamefont {Tannock}, \citenamefont
  {Löfwander},\ and\ \citenamefont {Kinaret}}]{Ferrari_2015}%
  \BibitemOpen
  \bibfield  {author} {\bibinfo {author} {\bibfnamefont {A.~C.}\ \bibnamefont
  {Ferrari}}, \bibinfo {author} {\bibfnamefont {F.}~\bibnamefont {Bonaccorso}},
  \bibinfo {author} {\bibfnamefont {V.}~\bibnamefont {Fal{'}ko}}, \bibinfo
  {author} {\bibfnamefont {K.~S.}\ \bibnamefont {Novoselov}}, \bibinfo {author}
  {\bibfnamefont {S.}~\bibnamefont {Roche}}, \bibinfo {author} {\bibfnamefont
  {P.}~\bibnamefont {Boggild}}, \bibinfo {author} {\bibfnamefont
  {S.}~\bibnamefont {Borini}}, \bibinfo {author} {\bibfnamefont {F.~H.~L.}\
  \bibnamefont {Koppens}}, \bibinfo {author} {\bibfnamefont {V.}~\bibnamefont
  {Palermo}}, \bibinfo {author} {\bibfnamefont {N.}~\bibnamefont {Pugno}},
  \bibinfo {author} {\bibfnamefont {J.~A.}\ \bibnamefont {Garrido}}, \bibinfo
  {author} {\bibfnamefont {R.}~\bibnamefont {Sordan}}, \bibinfo {author}
  {\bibfnamefont {A.}~\bibnamefont {Bianco}}, \bibinfo {author} {\bibfnamefont
  {L.}~\bibnamefont {Ballerini}}, \bibinfo {author} {\bibfnamefont
  {M.}~\bibnamefont {Prato}}, \bibinfo {author} {\bibfnamefont
  {E.}~\bibnamefont {Lidorikis}}, \bibinfo {author} {\bibfnamefont
  {J.}~\bibnamefont {Kivioja}}, \bibinfo {author} {\bibfnamefont
  {C.}~\bibnamefont {Marinelli}}, \bibinfo {author} {\bibfnamefont
  {T.}~\bibnamefont {Ryhänen}}, \bibinfo {author} {\bibfnamefont
  {A.}~\bibnamefont {Morpurgo}}, \bibinfo {author} {\bibfnamefont {J.~N.}\
  \bibnamefont {Coleman}}, \bibinfo {author} {\bibfnamefont {V.}~\bibnamefont
  {Nicolosi}}, \bibinfo {author} {\bibfnamefont {L.}~\bibnamefont {Colombo}},
  \bibinfo {author} {\bibfnamefont {A.}~\bibnamefont {Fert}}, \bibinfo {author}
  {\bibfnamefont {M.}~\bibnamefont {Garcia-Hernandez}}, \bibinfo {author}
  {\bibfnamefont {A.}~\bibnamefont {Bachtold}}, \bibinfo {author}
  {\bibfnamefont {G.~F.}\ \bibnamefont {Schneider}}, \bibinfo {author}
  {\bibfnamefont {F.}~\bibnamefont {Guinea}}, \bibinfo {author} {\bibfnamefont
  {C.}~\bibnamefont {Dekker}}, \bibinfo {author} {\bibfnamefont
  {M.}~\bibnamefont {Barbone}}, \bibinfo {author} {\bibfnamefont
  {Z.}~\bibnamefont {Sun}}, \bibinfo {author} {\bibfnamefont {C.}~\bibnamefont
  {Galiotis}}, \bibinfo {author} {\bibfnamefont {A.~N.}\ \bibnamefont
  {Grigorenko}}, \bibinfo {author} {\bibfnamefont {G.}~\bibnamefont
  {Konstantatos}}, \bibinfo {author} {\bibfnamefont {A.}~\bibnamefont {Kis}},
  \bibinfo {author} {\bibfnamefont {M.}~\bibnamefont {Katsnelson}}, \bibinfo
  {author} {\bibfnamefont {L.}~\bibnamefont {Vandersypen}}, \bibinfo {author}
  {\bibfnamefont {A.}~\bibnamefont {Loiseau}}, \bibinfo {author} {\bibfnamefont
  {V.}~\bibnamefont {Morandi}}, \bibinfo {author} {\bibfnamefont
  {D.}~\bibnamefont {Neumaier}}, \bibinfo {author} {\bibfnamefont
  {E.}~\bibnamefont {Treossi}}, \bibinfo {author} {\bibfnamefont
  {V.}~\bibnamefont {Pellegrini}}, \bibinfo {author} {\bibfnamefont
  {M.}~\bibnamefont {Polini}}, \bibinfo {author} {\bibfnamefont
  {A.}~\bibnamefont {Tredicucci}}, \bibinfo {author} {\bibfnamefont {G.~M.}\
  \bibnamefont {Williams}}, \bibinfo {author} {\bibfnamefont {B.}~\bibnamefont
  {Hee~Hong}}, \bibinfo {author} {\bibfnamefont {J.-H.}\ \bibnamefont {Ahn}},
  \bibinfo {author} {\bibfnamefont {J.}~\bibnamefont {Min~Kim}}, \bibinfo
  {author} {\bibfnamefont {H.}~\bibnamefont {Zirath}}, \bibinfo {author}
  {\bibfnamefont {B.~J.}\ \bibnamefont {van Wees}}, \bibinfo {author}
  {\bibfnamefont {H.}~\bibnamefont {van~der Zant}}, \bibinfo {author}
  {\bibfnamefont {L.}~\bibnamefont {Occhipinti}}, \bibinfo {author}
  {\bibfnamefont {A.}~\bibnamefont {Di~Matteo}}, \bibinfo {author}
  {\bibfnamefont {I.~A.}\ \bibnamefont {Kinloch}}, \bibinfo {author}
  {\bibfnamefont {T.}~\bibnamefont {Seyller}}, \bibinfo {author} {\bibfnamefont
  {E.}~\bibnamefont {Quesnel}}, \bibinfo {author} {\bibfnamefont
  {X.}~\bibnamefont {Feng}}, \bibinfo {author} {\bibfnamefont {K.}~\bibnamefont
  {Teo}}, \bibinfo {author} {\bibfnamefont {N.}~\bibnamefont {Rupesinghe}},
  \bibinfo {author} {\bibfnamefont {P.}~\bibnamefont {Hakonen}}, \bibinfo
  {author} {\bibfnamefont {S.~R.~T.}\ \bibnamefont {Neil}}, \bibinfo {author}
  {\bibfnamefont {Q.}~\bibnamefont {Tannock}}, \bibinfo {author} {\bibfnamefont
  {T.}~\bibnamefont {Löfwander}}, \ and\ \bibinfo {author} {\bibfnamefont
  {J.}~\bibnamefont {Kinaret}},\ }\href {\doibase 10.1039/C4NR01600A}
  {\bibfield  {journal} {\bibinfo  {journal} {Nanoscale}\ }\textbf {\bibinfo
  {volume} {7}},\ \bibinfo {pages} {4598} (\bibinfo {year} {2015})}\BibitemShut
  {NoStop}%
\bibitem [{\citenamefont {Bhimanapati}\ \emph {et~al.}(2015)\citenamefont
  {Bhimanapati}, \citenamefont {Lin}, \citenamefont {Meunier}, \citenamefont
  {Jung}, \citenamefont {Cha}, \citenamefont {Das}, \citenamefont {Xiao},
  \citenamefont {Son}, \citenamefont {Strano}, \citenamefont {Cooper},
  \citenamefont {Liang}, \citenamefont {Louie}, \citenamefont {Ringe},
  \citenamefont {Zhou}, \citenamefont {Kim}, \citenamefont {Naik},
  \citenamefont {Sumpter}, \citenamefont {Terrones}, \citenamefont {Xia},
  \citenamefont {Wang}, \citenamefont {Zhu}, \citenamefont {Akinwande},
  \citenamefont {Alem}, \citenamefont {Schuller}, \citenamefont {Schaak},
  \citenamefont {Terrones},\ and\ \citenamefont {Robinson}}]{Bhimanapati_2015}%
  \BibitemOpen
  \bibfield  {author} {\bibinfo {author} {\bibfnamefont {G.~R.}\ \bibnamefont
  {Bhimanapati}}, \bibinfo {author} {\bibfnamefont {Z.}~\bibnamefont {Lin}},
  \bibinfo {author} {\bibfnamefont {V.}~\bibnamefont {Meunier}}, \bibinfo
  {author} {\bibfnamefont {Y.}~\bibnamefont {Jung}}, \bibinfo {author}
  {\bibfnamefont {J.}~\bibnamefont {Cha}}, \bibinfo {author} {\bibfnamefont
  {S.}~\bibnamefont {Das}}, \bibinfo {author} {\bibfnamefont {D.}~\bibnamefont
  {Xiao}}, \bibinfo {author} {\bibfnamefont {Y.}~\bibnamefont {Son}}, \bibinfo
  {author} {\bibfnamefont {M.~S.}\ \bibnamefont {Strano}}, \bibinfo {author}
  {\bibfnamefont {V.~R.}\ \bibnamefont {Cooper}}, \bibinfo {author}
  {\bibfnamefont {L.}~\bibnamefont {Liang}}, \bibinfo {author} {\bibfnamefont
  {S.~G.}\ \bibnamefont {Louie}}, \bibinfo {author} {\bibfnamefont
  {E.}~\bibnamefont {Ringe}}, \bibinfo {author} {\bibfnamefont
  {W.}~\bibnamefont {Zhou}}, \bibinfo {author} {\bibfnamefont {S.~S.}\
  \bibnamefont {Kim}}, \bibinfo {author} {\bibfnamefont {R.~R.}\ \bibnamefont
  {Naik}}, \bibinfo {author} {\bibfnamefont {B.~G.}\ \bibnamefont {Sumpter}},
  \bibinfo {author} {\bibfnamefont {H.}~\bibnamefont {Terrones}}, \bibinfo
  {author} {\bibfnamefont {F.}~\bibnamefont {Xia}}, \bibinfo {author}
  {\bibfnamefont {Y.}~\bibnamefont {Wang}}, \bibinfo {author} {\bibfnamefont
  {J.}~\bibnamefont {Zhu}}, \bibinfo {author} {\bibfnamefont {D.}~\bibnamefont
  {Akinwande}}, \bibinfo {author} {\bibfnamefont {N.}~\bibnamefont {Alem}},
  \bibinfo {author} {\bibfnamefont {J.~A.}\ \bibnamefont {Schuller}}, \bibinfo
  {author} {\bibfnamefont {R.~E.}\ \bibnamefont {Schaak}}, \bibinfo {author}
  {\bibfnamefont {M.}~\bibnamefont {Terrones}}, \ and\ \bibinfo {author}
  {\bibfnamefont {J.~A.}\ \bibnamefont {Robinson}},\ }\href {\doibase
  10.1021/acsnano.5b05556} {\bibfield  {journal} {\bibinfo  {journal} {ACS
  Nano}\ }\textbf {\bibinfo {volume} {9}},\ \bibinfo {pages} {11509} (\bibinfo
  {year} {2015})}\BibitemShut {NoStop}%
\bibitem [{\citenamefont {Trevisanutto}\ \emph {et~al.}(2008)\citenamefont
  {Trevisanutto}, \citenamefont {Giorgetti}, \citenamefont {Reining},
  \citenamefont {Ladisa},\ and\ \citenamefont {Olevano}}]{trevisanutto2008ab}%
  \BibitemOpen
  \bibfield  {author} {\bibinfo {author} {\bibfnamefont {P.~E.}\ \bibnamefont
  {Trevisanutto}}, \bibinfo {author} {\bibfnamefont {C.}~\bibnamefont
  {Giorgetti}}, \bibinfo {author} {\bibfnamefont {L.}~\bibnamefont {Reining}},
  \bibinfo {author} {\bibfnamefont {M.}~\bibnamefont {Ladisa}}, \ and\ \bibinfo
  {author} {\bibfnamefont {V.}~\bibnamefont {Olevano}},\ }\href {\doibase
  10.1103/PhysRevLett.101.226405} {\bibfield  {journal} {\bibinfo  {journal}
  {Phys. Rev. Lett.}\ }\textbf {\bibinfo {volume} {101}},\ \bibinfo {pages}
  {226405} (\bibinfo {year} {2008})}\BibitemShut {NoStop}%
\bibitem [{\citenamefont {Haastrup}\ \emph {et~al.}(2018)\citenamefont
  {Haastrup}, \citenamefont {Strange}, \citenamefont {Pandey}, \citenamefont
  {Deilmann}, \citenamefont {Schmidt}, \citenamefont {Hinsche}, \citenamefont
  {Gjerding}, \citenamefont {Torelli}, \citenamefont {Larsen}, \citenamefont
  {Riis-Jensen} \emph {et~al.}}]{haastrup2018computational}%
  \BibitemOpen
  \bibfield  {author} {\bibinfo {author} {\bibfnamefont {S.}~\bibnamefont
  {Haastrup}}, \bibinfo {author} {\bibfnamefont {M.}~\bibnamefont {Strange}},
  \bibinfo {author} {\bibfnamefont {M.}~\bibnamefont {Pandey}}, \bibinfo
  {author} {\bibfnamefont {T.}~\bibnamefont {Deilmann}}, \bibinfo {author}
  {\bibfnamefont {P.~S.}\ \bibnamefont {Schmidt}}, \bibinfo {author}
  {\bibfnamefont {N.~F.}\ \bibnamefont {Hinsche}}, \bibinfo {author}
  {\bibfnamefont {M.~N.}\ \bibnamefont {Gjerding}}, \bibinfo {author}
  {\bibfnamefont {D.}~\bibnamefont {Torelli}}, \bibinfo {author} {\bibfnamefont
  {P.~M.}\ \bibnamefont {Larsen}}, \bibinfo {author} {\bibfnamefont {A.~C.}\
  \bibnamefont {Riis-Jensen}},  \emph {et~al.},\ }\href {\doibase
  10.1088/2053-1583/aacfc1} {\bibfield  {journal} {\bibinfo  {journal} {2D
  Materials}\ }\textbf {\bibinfo {volume} {5}},\ \bibinfo {pages} {042002}
  (\bibinfo {year} {2018})}\BibitemShut {NoStop}%
\bibitem [{\citenamefont {Rasmussen}\ \emph {et~al.}(2021)\citenamefont
  {Rasmussen}, \citenamefont {Deilmann},\ and\ \citenamefont
  {Thygesen}}]{rasmussen2021towards}%
  \BibitemOpen
  \bibfield  {author} {\bibinfo {author} {\bibfnamefont {A.}~\bibnamefont
  {Rasmussen}}, \bibinfo {author} {\bibfnamefont {T.}~\bibnamefont {Deilmann}},
  \ and\ \bibinfo {author} {\bibfnamefont {K.~S.}\ \bibnamefont {Thygesen}},\
  }\href {\doibase 10.1038/s41524-020-00480-7} {\bibfield  {journal} {\bibinfo
  {journal} {npj Comput. Materials}\ }\textbf {\bibinfo {volume} {7}},\
  \bibinfo {pages} {1} (\bibinfo {year} {2021})}\BibitemShut {NoStop}%
\bibitem [{\citenamefont {Qian}\ \emph {et~al.}(2014)\citenamefont {Qian},
  \citenamefont {Liu}, \citenamefont {Fu},\ and\ \citenamefont
  {Li}}]{qian2014quantum}%
  \BibitemOpen
  \bibfield  {author} {\bibinfo {author} {\bibfnamefont {X.}~\bibnamefont
  {Qian}}, \bibinfo {author} {\bibfnamefont {J.}~\bibnamefont {Liu}}, \bibinfo
  {author} {\bibfnamefont {L.}~\bibnamefont {Fu}}, \ and\ \bibinfo {author}
  {\bibfnamefont {J.}~\bibnamefont {Li}},\ }\href {\doibase
  10.1126/science.1256815} {\bibfield  {journal} {\bibinfo  {journal}
  {Science}\ }\textbf {\bibinfo {volume} {346}},\ \bibinfo {pages} {1344}
  (\bibinfo {year} {2014})}\BibitemShut {NoStop}%
\bibitem [{\citenamefont {Varsano}\ \emph {et~al.}(2020)\citenamefont
  {Varsano}, \citenamefont {Palummo}, \citenamefont {Molinari},\ and\
  \citenamefont {Rontani}}]{varsano2020monolayer}%
  \BibitemOpen
  \bibfield  {author} {\bibinfo {author} {\bibfnamefont {D.}~\bibnamefont
  {Varsano}}, \bibinfo {author} {\bibfnamefont {M.}~\bibnamefont {Palummo}},
  \bibinfo {author} {\bibfnamefont {E.}~\bibnamefont {Molinari}}, \ and\
  \bibinfo {author} {\bibfnamefont {M.}~\bibnamefont {Rontani}},\ }\href
  {\doibase 10.1038/s41565-020-0650-4} {\bibfield  {journal} {\bibinfo
  {journal} {Nature Nanotech.}\ }\textbf {\bibinfo {volume} {15}},\ \bibinfo
  {pages} {367} (\bibinfo {year} {2020})}\BibitemShut {NoStop}%
\bibitem [{\citenamefont {Qiu}\ \emph {et~al.}(2016{\natexlab{a}})\citenamefont
  {Qiu}, \citenamefont {Felipe},\ and\ \citenamefont
  {Louie}}]{qiu2016screening}%
  \BibitemOpen
  \bibfield  {author} {\bibinfo {author} {\bibfnamefont {D.~Y.}\ \bibnamefont
  {Qiu}}, \bibinfo {author} {\bibfnamefont {H.}~\bibnamefont {Felipe}}, \ and\
  \bibinfo {author} {\bibfnamefont {S.~G.}\ \bibnamefont {Louie}},\ }\href
  {\doibase 10.1103/PhysRevB.93.235435} {\bibfield  {journal} {\bibinfo
  {journal} {Phys. Rev. B}\ }\textbf {\bibinfo {volume} {93}},\ \bibinfo
  {pages} {235435} (\bibinfo {year} {2016}{\natexlab{a}})}\BibitemShut
  {NoStop}%
\bibitem [{\citenamefont {Qiu}\ \emph {et~al.}(2017)\citenamefont {Qiu},
  \citenamefont {da~Jornada},\ and\ \citenamefont {Louie}}]{Qiu_2017}%
  \BibitemOpen
  \bibfield  {author} {\bibinfo {author} {\bibfnamefont {D.~Y.}\ \bibnamefont
  {Qiu}}, \bibinfo {author} {\bibfnamefont {F.~H.}\ \bibnamefont {da~Jornada}},
  \ and\ \bibinfo {author} {\bibfnamefont {S.~G.}\ \bibnamefont {Louie}},\
  }\href {\doibase 10.1021/acs.nanolett.7b01365} {\bibfield  {journal}
  {\bibinfo  {journal} {Nano Letters}\ }\textbf {\bibinfo {volume} {17}},\
  \bibinfo {pages} {4706} (\bibinfo {year} {2017})}\BibitemShut {NoStop}%
\bibitem [{\citenamefont {da~Jornada}\ \emph {et~al.}(2020)\citenamefont
  {da~Jornada}, \citenamefont {Xian}, \citenamefont {Rubio},\ and\
  \citenamefont {Louie}}]{DaJornada_2020}%
  \BibitemOpen
  \bibfield  {author} {\bibinfo {author} {\bibfnamefont {F.~H.}\ \bibnamefont
  {da~Jornada}}, \bibinfo {author} {\bibfnamefont {L.}~\bibnamefont {Xian}},
  \bibinfo {author} {\bibfnamefont {A.}~\bibnamefont {Rubio}}, \ and\ \bibinfo
  {author} {\bibfnamefont {S.~G.}\ \bibnamefont {Louie}},\ }\href@noop {}
  {\bibfield  {journal} {\bibinfo  {journal} {Nature communications}\ }\textbf
  {\bibinfo {volume} {11}},\ \bibinfo {pages} {1} (\bibinfo {year}
  {2020})}\BibitemShut {NoStop}%
\bibitem [{\citenamefont {Wu}\ \emph {et~al.}(2018)\citenamefont {Wu},
  \citenamefont {Xia}, \citenamefont {Gao}, \citenamefont {Jia}, \citenamefont
  {Zhang},\ and\ \citenamefont {Ren}}]{Yabei_2018}%
  \BibitemOpen
  \bibfield  {author} {\bibinfo {author} {\bibfnamefont {Y.}~\bibnamefont
  {Wu}}, \bibinfo {author} {\bibfnamefont {W.}~\bibnamefont {Xia}}, \bibinfo
  {author} {\bibfnamefont {W.}~\bibnamefont {Gao}}, \bibinfo {author}
  {\bibfnamefont {F.}~\bibnamefont {Jia}}, \bibinfo {author} {\bibfnamefont
  {P.}~\bibnamefont {Zhang}}, \ and\ \bibinfo {author} {\bibfnamefont
  {W.}~\bibnamefont {Ren}},\ }\href@noop {} {\bibfield  {journal} {\bibinfo
  {journal} {2D Materials}\ }\textbf {\bibinfo {volume} {6}},\ \bibinfo {pages}
  {015018} (\bibinfo {year} {2018})}\BibitemShut {NoStop}%
\bibitem [{\citenamefont {Zhang}\ \emph {et~al.}(2019)\citenamefont {Zhang},
  \citenamefont {Xia}, \citenamefont {Wu},\ and\ \citenamefont
  {Zhang}}]{Zhang_2019}%
  \BibitemOpen
  \bibfield  {author} {\bibinfo {author} {\bibfnamefont {Y.}~\bibnamefont
  {Zhang}}, \bibinfo {author} {\bibfnamefont {W.}~\bibnamefont {Xia}}, \bibinfo
  {author} {\bibfnamefont {Y.}~\bibnamefont {Wu}}, \ and\ \bibinfo {author}
  {\bibfnamefont {P.}~\bibnamefont {Zhang}},\ }\href {\doibase
  10.1039/C9NR01160A} {\bibfield  {journal} {\bibinfo  {journal} {Nanoscale}\
  }\textbf {\bibinfo {volume} {11}},\ \bibinfo {pages} {3993} (\bibinfo {year}
  {2019})}\BibitemShut {NoStop}%
\bibitem [{\citenamefont {Qiu}\ \emph {et~al.}(2020)\citenamefont {Qiu},
  \citenamefont {Coh}, \citenamefont {Cohen},\ and\ \citenamefont
  {Louie}}]{Qiu_2020}%
  \BibitemOpen
  \bibfield  {author} {\bibinfo {author} {\bibfnamefont {D.~Y.}\ \bibnamefont
  {Qiu}}, \bibinfo {author} {\bibfnamefont {S.}~\bibnamefont {Coh}}, \bibinfo
  {author} {\bibfnamefont {M.~L.}\ \bibnamefont {Cohen}}, \ and\ \bibinfo
  {author} {\bibfnamefont {S.~G.}\ \bibnamefont {Louie}},\ }\href {\doibase
  10.1103/PhysRevB.101.235154} {\bibfield  {journal} {\bibinfo  {journal}
  {Phys. Rev. B}\ }\textbf {\bibinfo {volume} {101}},\ \bibinfo {pages}
  {235154} (\bibinfo {year} {2020})}\BibitemShut {NoStop}%
\bibitem [{\citenamefont {Ismail-Beigi}(2006)}]{Beigi_2006}%
  \BibitemOpen
  \bibfield  {author} {\bibinfo {author} {\bibfnamefont {S.}~\bibnamefont
  {Ismail-Beigi}},\ }\href {\doibase 10.1103/PhysRevB.73.233103} {\bibfield
  {journal} {\bibinfo  {journal} {Phys. Rev. B}\ }\textbf {\bibinfo {volume}
  {73}},\ \bibinfo {pages} {233103} (\bibinfo {year} {2006})}\BibitemShut
  {NoStop}%
\bibitem [{\citenamefont {Rozzi}\ \emph {et~al.}(2006)\citenamefont {Rozzi},
  \citenamefont {Varsano}, \citenamefont {Marini}, \citenamefont {Gross},\ and\
  \citenamefont {Rubio}}]{Rozzi_2006}%
  \BibitemOpen
  \bibfield  {author} {\bibinfo {author} {\bibfnamefont {C.~A.}\ \bibnamefont
  {Rozzi}}, \bibinfo {author} {\bibfnamefont {D.}~\bibnamefont {Varsano}},
  \bibinfo {author} {\bibfnamefont {A.}~\bibnamefont {Marini}}, \bibinfo
  {author} {\bibfnamefont {E.~K.~U.}\ \bibnamefont {Gross}}, \ and\ \bibinfo
  {author} {\bibfnamefont {A.}~\bibnamefont {Rubio}},\ }\href {\doibase
  10.1103/PhysRevB.73.205119} {\bibfield  {journal} {\bibinfo  {journal} {Phys.
  Rev. B}\ }\textbf {\bibinfo {volume} {73}},\ \bibinfo {pages} {205119}
  (\bibinfo {year} {2006})}\BibitemShut {NoStop}%
\bibitem [{\citenamefont {Cudazzo}\ \emph {et~al.}(2011)\citenamefont
  {Cudazzo}, \citenamefont {Tokatly},\ and\ \citenamefont
  {Rubio}}]{Cudazzo_2011}%
  \BibitemOpen
  \bibfield  {author} {\bibinfo {author} {\bibfnamefont {P.}~\bibnamefont
  {Cudazzo}}, \bibinfo {author} {\bibfnamefont {I.~V.}\ \bibnamefont
  {Tokatly}}, \ and\ \bibinfo {author} {\bibfnamefont {A.}~\bibnamefont
  {Rubio}},\ }\href {\doibase 10.1103/PhysRevB.84.085406} {\bibfield  {journal}
  {\bibinfo  {journal} {Phys. Rev. B}\ }\textbf {\bibinfo {volume} {84}},\
  \bibinfo {pages} {085406} (\bibinfo {year} {2011})}\BibitemShut {NoStop}%
\bibitem [{\citenamefont {H\"user}\ \emph {et~al.}(2013)\citenamefont
  {H\"user}, \citenamefont {Olsen},\ and\ \citenamefont
  {Thygesen}}]{Huser_2013}%
  \BibitemOpen
  \bibfield  {author} {\bibinfo {author} {\bibfnamefont {F.}~\bibnamefont
  {H\"user}}, \bibinfo {author} {\bibfnamefont {T.}~\bibnamefont {Olsen}}, \
  and\ \bibinfo {author} {\bibfnamefont {K.~S.}\ \bibnamefont {Thygesen}},\
  }\href {\doibase 10.1103/PhysRevB.88.245309} {\bibfield  {journal} {\bibinfo
  {journal} {Phys. Rev. B}\ }\textbf {\bibinfo {volume} {88}},\ \bibinfo
  {pages} {245309} (\bibinfo {year} {2013})}\BibitemShut {NoStop}%
\bibitem [{\citenamefont {Thygesen}(2017)}]{Thygesen_2017}%
  \BibitemOpen
  \bibfield  {author} {\bibinfo {author} {\bibfnamefont {K.~S.}\ \bibnamefont
  {Thygesen}},\ }\href {\doibase 10.1088/2053-1583/aa6432} {\bibfield
  {journal} {\bibinfo  {journal} {2D Materials}\ }\textbf {\bibinfo {volume}
  {4}},\ \bibinfo {pages} {022004} (\bibinfo {year} {2017})}\BibitemShut
  {NoStop}%
\bibitem [{\citenamefont {H{\"u}ser}\ \emph {et~al.}(2013)\citenamefont
  {H{\"u}ser}, \citenamefont {Olsen},\ and\ \citenamefont
  {Thygesen}}]{huser2013dielectric}%
  \BibitemOpen
  \bibfield  {author} {\bibinfo {author} {\bibfnamefont {F.}~\bibnamefont
  {H{\"u}ser}}, \bibinfo {author} {\bibfnamefont {T.}~\bibnamefont {Olsen}}, \
  and\ \bibinfo {author} {\bibfnamefont {K.~S.}\ \bibnamefont {Thygesen}},\
  }\href {\doibase 10.1103/PhysRevB.88.245309} {\bibfield  {journal} {\bibinfo
  {journal} {Phys. Rev. B}\ }\textbf {\bibinfo {volume} {88}},\ \bibinfo
  {pages} {245309} (\bibinfo {year} {2013})}\BibitemShut {NoStop}%
\bibitem [{\citenamefont {Qiu}\ \emph {et~al.}(2016{\natexlab{b}})\citenamefont
  {Qiu}, \citenamefont {da~Jornada},\ and\ \citenamefont {Louie}}]{Qiu_2016}%
  \BibitemOpen
  \bibfield  {author} {\bibinfo {author} {\bibfnamefont {D.~Y.}\ \bibnamefont
  {Qiu}}, \bibinfo {author} {\bibfnamefont {F.~H.}\ \bibnamefont {da~Jornada}},
  \ and\ \bibinfo {author} {\bibfnamefont {S.~G.}\ \bibnamefont {Louie}},\
  }\href {\doibase 10.1103/PhysRevB.93.235435} {\bibfield  {journal} {\bibinfo
  {journal} {Phys. Rev. B}\ }\textbf {\bibinfo {volume} {93}},\ \bibinfo
  {pages} {235435} (\bibinfo {year} {2016}{\natexlab{b}})}\BibitemShut
  {NoStop}%
\bibitem [{\citenamefont {Rasmussen}\ \emph {et~al.}(2016)\citenamefont
  {Rasmussen}, \citenamefont {Schmidt}, \citenamefont {Winther},\ and\
  \citenamefont {Thygesen}}]{Rasmussen_2016}%
  \BibitemOpen
  \bibfield  {author} {\bibinfo {author} {\bibfnamefont {F.~A.}\ \bibnamefont
  {Rasmussen}}, \bibinfo {author} {\bibfnamefont {P.~S.}\ \bibnamefont
  {Schmidt}}, \bibinfo {author} {\bibfnamefont {K.~T.}\ \bibnamefont
  {Winther}}, \ and\ \bibinfo {author} {\bibfnamefont {K.~S.}\ \bibnamefont
  {Thygesen}},\ }\href {\doibase 10.1103/PhysRevB.94.155406} {\bibfield
  {journal} {\bibinfo  {journal} {Phys. Rev. B}\ }\textbf {\bibinfo {volume}
  {94}},\ \bibinfo {pages} {155406} (\bibinfo {year} {2016})}\BibitemShut
  {NoStop}%
\bibitem [{\citenamefont {da~Jornada}\ \emph {et~al.}(2017)\citenamefont
  {da~Jornada}, \citenamefont {Qiu},\ and\ \citenamefont
  {Louie}}]{daJornada_2017}%
  \BibitemOpen
  \bibfield  {author} {\bibinfo {author} {\bibfnamefont {F.~H.}\ \bibnamefont
  {da~Jornada}}, \bibinfo {author} {\bibfnamefont {D.~Y.}\ \bibnamefont {Qiu}},
  \ and\ \bibinfo {author} {\bibfnamefont {S.~G.}\ \bibnamefont {Louie}},\
  }\href {\doibase 10.1103/PhysRevB.95.035109} {\bibfield  {journal} {\bibinfo
  {journal} {Phys. Rev. B}\ }\textbf {\bibinfo {volume} {95}},\ \bibinfo
  {pages} {035109} (\bibinfo {year} {2017})}\BibitemShut {NoStop}%
\bibitem [{\citenamefont {Chernikov}\ \emph {et~al.}(2014)\citenamefont
  {Chernikov}, \citenamefont {Berkelbach}, \citenamefont {Hill}, \citenamefont
  {Rigosi}, \citenamefont {Li}, \citenamefont {Aslan}, \citenamefont
  {Reichman}, \citenamefont {Hybertsen},\ and\ \citenamefont
  {Heinz}}]{Chernikov_2014}%
  \BibitemOpen
  \bibfield  {author} {\bibinfo {author} {\bibfnamefont {A.}~\bibnamefont
  {Chernikov}}, \bibinfo {author} {\bibfnamefont {T.~C.}\ \bibnamefont
  {Berkelbach}}, \bibinfo {author} {\bibfnamefont {H.~M.}\ \bibnamefont
  {Hill}}, \bibinfo {author} {\bibfnamefont {A.}~\bibnamefont {Rigosi}},
  \bibinfo {author} {\bibfnamefont {Y.}~\bibnamefont {Li}}, \bibinfo {author}
  {\bibfnamefont {O.~B.}\ \bibnamefont {Aslan}}, \bibinfo {author}
  {\bibfnamefont {D.~R.}\ \bibnamefont {Reichman}}, \bibinfo {author}
  {\bibfnamefont {M.~S.}\ \bibnamefont {Hybertsen}}, \ and\ \bibinfo {author}
  {\bibfnamefont {T.~F.}\ \bibnamefont {Heinz}},\ }\href {\doibase
  10.1103/PhysRevLett.113.076802} {\bibfield  {journal} {\bibinfo  {journal}
  {Phys. Rev. Lett.}\ }\textbf {\bibinfo {volume} {113}},\ \bibinfo {pages}
  {076802} (\bibinfo {year} {2014})}\BibitemShut {NoStop}%
\bibitem [{\citenamefont {Xia}\ \emph {et~al.}(2020)\citenamefont {Xia},
  \citenamefont {Gao}, \citenamefont {Lopez-Candales}, \citenamefont {Wu},
  \citenamefont {Ren}, \citenamefont {Zhang},\ and\ \citenamefont
  {Zhang}}]{Xia_2020}%
  \BibitemOpen
  \bibfield  {author} {\bibinfo {author} {\bibfnamefont {W.}~\bibnamefont
  {Xia}}, \bibinfo {author} {\bibfnamefont {W.}~\bibnamefont {Gao}}, \bibinfo
  {author} {\bibfnamefont {G.}~\bibnamefont {Lopez-Candales}}, \bibinfo
  {author} {\bibfnamefont {Y.}~\bibnamefont {Wu}}, \bibinfo {author}
  {\bibfnamefont {W.}~\bibnamefont {Ren}}, \bibinfo {author} {\bibfnamefont
  {W.}~\bibnamefont {Zhang}}, \ and\ \bibinfo {author} {\bibfnamefont
  {P.}~\bibnamefont {Zhang}},\ }\href {\doibase 10.1038/s41524-020-00385-5}
  {\bibfield  {journal} {\bibinfo  {journal} {npj Computational Materials}\
  }\textbf {\bibinfo {volume} {6}},\ \bibinfo {pages} {118} (\bibinfo {year}
  {2020})}\BibitemShut {NoStop}%
\bibitem [{\citenamefont {Marini}\ \emph {et~al.}(2009)\citenamefont {Marini},
  \citenamefont {Hogan}, \citenamefont {Grüning},\ and\ \citenamefont
  {Varsano}}]{yambo_2009}%
  \BibitemOpen
  \bibfield  {author} {\bibinfo {author} {\bibfnamefont {A.}~\bibnamefont
  {Marini}}, \bibinfo {author} {\bibfnamefont {C.}~\bibnamefont {Hogan}},
  \bibinfo {author} {\bibfnamefont {M.}~\bibnamefont {Grüning}}, \ and\
  \bibinfo {author} {\bibfnamefont {D.}~\bibnamefont {Varsano}},\ }\href
  {\doibase https://doi.org/10.1016/j.cpc.2009.02.003} {\bibfield  {journal}
  {\bibinfo  {journal} {Comput. Phys. Commun.}\ }\textbf {\bibinfo {volume}
  {180}},\ \bibinfo {pages} {1392 } (\bibinfo {year} {2009})}\BibitemShut
  {NoStop}%
\bibitem [{\citenamefont {Sangalli}\ \emph {et~al.}(2019)\citenamefont
  {Sangalli}, \citenamefont {Ferretti}, \citenamefont {Miranda}, \citenamefont
  {Attaccalite}, \citenamefont {Marri}, \citenamefont {Cannuccia},
  \citenamefont {Melo}, \citenamefont {Marsili}, \citenamefont {Paleari},
  \citenamefont {Marrazzo}, \citenamefont {Prandini}, \citenamefont
  {Bonf{\`{a}}}, \citenamefont {Atambo}, \citenamefont {Affinito},
  \citenamefont {Palummo}, \citenamefont {Molina-S{\'{a}}nchez}, \citenamefont
  {Hogan}, \citenamefont {Grüning}, \citenamefont {Varsano},\ and\
  \citenamefont {Marini}}]{yambo_2019}%
  \BibitemOpen
  \bibfield  {author} {\bibinfo {author} {\bibfnamefont {D.}~\bibnamefont
  {Sangalli}}, \bibinfo {author} {\bibfnamefont {A.}~\bibnamefont {Ferretti}},
  \bibinfo {author} {\bibfnamefont {H.}~\bibnamefont {Miranda}}, \bibinfo
  {author} {\bibfnamefont {C.}~\bibnamefont {Attaccalite}}, \bibinfo {author}
  {\bibfnamefont {I.}~\bibnamefont {Marri}}, \bibinfo {author} {\bibfnamefont
  {E.}~\bibnamefont {Cannuccia}}, \bibinfo {author} {\bibfnamefont
  {P.}~\bibnamefont {Melo}}, \bibinfo {author} {\bibfnamefont {M.}~\bibnamefont
  {Marsili}}, \bibinfo {author} {\bibfnamefont {F.}~\bibnamefont {Paleari}},
  \bibinfo {author} {\bibfnamefont {A.}~\bibnamefont {Marrazzo}}, \bibinfo
  {author} {\bibfnamefont {G.}~\bibnamefont {Prandini}}, \bibinfo {author}
  {\bibfnamefont {P.}~\bibnamefont {Bonf{\`{a}}}}, \bibinfo {author}
  {\bibfnamefont {M.~O.}\ \bibnamefont {Atambo}}, \bibinfo {author}
  {\bibfnamefont {F.}~\bibnamefont {Affinito}}, \bibinfo {author}
  {\bibfnamefont {M.}~\bibnamefont {Palummo}}, \bibinfo {author} {\bibfnamefont
  {A.}~\bibnamefont {Molina-S{\'{a}}nchez}}, \bibinfo {author} {\bibfnamefont
  {C.}~\bibnamefont {Hogan}}, \bibinfo {author} {\bibfnamefont
  {M.}~\bibnamefont {Grüning}}, \bibinfo {author} {\bibfnamefont
  {D.}~\bibnamefont {Varsano}}, \ and\ \bibinfo {author} {\bibfnamefont
  {A.}~\bibnamefont {Marini}},\ }\href {\doibase 10.1088/1361-648x/ab15d0}
  {\bibfield  {journal} {\bibinfo  {journal} {J. Phys.: Condens. Matter}\
  }\textbf {\bibinfo {volume} {31}},\ \bibinfo {pages} {325902} (\bibinfo
  {year} {2019})}\BibitemShut {NoStop}%
\bibitem [{\citenamefont {Monkhorst}\ and\ \citenamefont
  {Pack}(1976)}]{Monkhorst_1976}%
  \BibitemOpen
  \bibfield  {author} {\bibinfo {author} {\bibfnamefont {H.~J.}\ \bibnamefont
  {Monkhorst}}\ and\ \bibinfo {author} {\bibfnamefont {J.~D.}\ \bibnamefont
  {Pack}},\ }\href {\doibase 10.1103/PhysRevB.13.5188} {\bibfield  {journal}
  {\bibinfo  {journal} {Phys. Rev. B}\ }\textbf {\bibinfo {volume} {13}},\
  \bibinfo {pages} {5188} (\bibinfo {year} {1976})}\BibitemShut {NoStop}%
\bibitem [{\citenamefont {Gygi}\ and\ \citenamefont
  {Baldereschi}(1986)}]{Gygi_1986}%
  \BibitemOpen
  \bibfield  {author} {\bibinfo {author} {\bibfnamefont {F.}~\bibnamefont
  {Gygi}}\ and\ \bibinfo {author} {\bibfnamefont {A.}~\bibnamefont
  {Baldereschi}},\ }\href {\doibase 10.1103/PhysRevB.34.4405} {\bibfield
  {journal} {\bibinfo  {journal} {Phys. Rev. B}\ }\textbf {\bibinfo {volume}
  {34}},\ \bibinfo {pages} {4405} (\bibinfo {year} {1986})}\BibitemShut
  {NoStop}%
\bibitem [{\citenamefont {Carrier}\ \emph {et~al.}(2007)\citenamefont
  {Carrier}, \citenamefont {Rohra},\ and\ \citenamefont
  {G{\"o}rling}}]{carrier2007general}%
  \BibitemOpen
  \bibfield  {author} {\bibinfo {author} {\bibfnamefont {P.}~\bibnamefont
  {Carrier}}, \bibinfo {author} {\bibfnamefont {S.}~\bibnamefont {Rohra}}, \
  and\ \bibinfo {author} {\bibfnamefont {A.}~\bibnamefont {G{\"o}rling}},\
  }\href {\doibase 10.1103/PhysRevB.75.205126} {\bibfield  {journal} {\bibinfo
  {journal} {Phys. Rev. B}\ }\textbf {\bibinfo {volume} {75}},\ \bibinfo
  {pages} {205126} (\bibinfo {year} {2007})}\BibitemShut {NoStop}%
\bibitem [{\citenamefont {Deslippe}\ \emph {et~al.}(2012)\citenamefont
  {Deslippe}, \citenamefont {Samsonidze}, \citenamefont {Strubbe},
  \citenamefont {Jain}, \citenamefont {Cohen},\ and\ \citenamefont
  {Louie}}]{deslippe2012berkeleygw}%
  \BibitemOpen
  \bibfield  {author} {\bibinfo {author} {\bibfnamefont {J.}~\bibnamefont
  {Deslippe}}, \bibinfo {author} {\bibfnamefont {G.}~\bibnamefont
  {Samsonidze}}, \bibinfo {author} {\bibfnamefont {D.~A.}\ \bibnamefont
  {Strubbe}}, \bibinfo {author} {\bibfnamefont {M.}~\bibnamefont {Jain}},
  \bibinfo {author} {\bibfnamefont {M.~L.}\ \bibnamefont {Cohen}}, \ and\
  \bibinfo {author} {\bibfnamefont {S.~G.}\ \bibnamefont {Louie}},\ }\href
  {\doibase 10.1016/j.cpc.2011.12.006} {\bibfield  {journal} {\bibinfo
  {journal} {Comput. Phys. Commun.}\ }\textbf {\bibinfo {volume} {183}},\
  \bibinfo {pages} {1269} (\bibinfo {year} {2012})}\BibitemShut {NoStop}%
\bibitem [{\citenamefont {Pulci}\ \emph {et~al.}(1998)\citenamefont {Pulci},
  \citenamefont {Onida}, \citenamefont {Del~Sole},\ and\ \citenamefont
  {Reining}}]{pulci1998ab}%
  \BibitemOpen
  \bibfield  {author} {\bibinfo {author} {\bibfnamefont {O.}~\bibnamefont
  {Pulci}}, \bibinfo {author} {\bibfnamefont {G.}~\bibnamefont {Onida}},
  \bibinfo {author} {\bibfnamefont {R.}~\bibnamefont {Del~Sole}}, \ and\
  \bibinfo {author} {\bibfnamefont {L.}~\bibnamefont {Reining}},\ }\href
  {\doibase 10.1103/PhysRevLett.81.5374} {\bibfield  {journal} {\bibinfo
  {journal} {Phys. Rev. Lett.}\ }\textbf {\bibinfo {volume} {81}},\ \bibinfo
  {pages} {5374} (\bibinfo {year} {1998})}\BibitemShut {NoStop}%
\bibitem [{\citenamefont {Rangel}\ \emph {et~al.}(2020)\citenamefont {Rangel},
  \citenamefont {Del~Ben}, \citenamefont {Varsano}, \citenamefont {Antonius},
  \citenamefont {Bruneval}, \citenamefont {Felipe}, \citenamefont {van Setten},
  \citenamefont {Orhan}, \citenamefont {O'Regan}, \citenamefont {Canning} \emph
  {et~al.}}]{rangel2020reproducibility}%
  \BibitemOpen
  \bibfield  {author} {\bibinfo {author} {\bibfnamefont {T.}~\bibnamefont
  {Rangel}}, \bibinfo {author} {\bibfnamefont {M.}~\bibnamefont {Del~Ben}},
  \bibinfo {author} {\bibfnamefont {D.}~\bibnamefont {Varsano}}, \bibinfo
  {author} {\bibfnamefont {G.}~\bibnamefont {Antonius}}, \bibinfo {author}
  {\bibfnamefont {F.}~\bibnamefont {Bruneval}}, \bibinfo {author}
  {\bibfnamefont {H.}~\bibnamefont {Felipe}}, \bibinfo {author} {\bibfnamefont
  {M.~J.}\ \bibnamefont {van Setten}}, \bibinfo {author} {\bibfnamefont
  {O.~K.}\ \bibnamefont {Orhan}}, \bibinfo {author} {\bibfnamefont {D.~D.}\
  \bibnamefont {O'Regan}}, \bibinfo {author} {\bibfnamefont {A.}~\bibnamefont
  {Canning}},  \emph {et~al.},\ }\href {\doibase 10.1016/j.cpc.2020.107242}
  {\bibfield  {journal} {\bibinfo  {journal} {Comput. Phys. Commun.}\ }\textbf
  {\bibinfo {volume} {255}},\ \bibinfo {pages} {107242} (\bibinfo {year}
  {2020})}\BibitemShut {NoStop}%
\bibitem [{\citenamefont {Godby}\ and\ \citenamefont
  {Needs}(1989)}]{Godby_1989}%
  \BibitemOpen
  \bibfield  {author} {\bibinfo {author} {\bibfnamefont {R.~W.}\ \bibnamefont
  {Godby}}\ and\ \bibinfo {author} {\bibfnamefont {R.~J.}\ \bibnamefont
  {Needs}},\ }\href {\doibase 10.1103/PhysRevLett.62.1169} {\bibfield
  {journal} {\bibinfo  {journal} {Phys. Rev. Lett.}\ }\textbf {\bibinfo
  {volume} {62}},\ \bibinfo {pages} {1169} (\bibinfo {year}
  {1989})}\BibitemShut {NoStop}%
\bibitem [{\citenamefont {Pick}\ \emph {et~al.}(1970)\citenamefont {Pick},
  \citenamefont {Cohen},\ and\ \citenamefont {Martin}}]{Pick_1970}%
  \BibitemOpen
  \bibfield  {author} {\bibinfo {author} {\bibfnamefont {R.~M.}\ \bibnamefont
  {Pick}}, \bibinfo {author} {\bibfnamefont {M.~H.}\ \bibnamefont {Cohen}}, \
  and\ \bibinfo {author} {\bibfnamefont {R.~M.}\ \bibnamefont {Martin}},\
  }\href {\doibase 10.1103/PhysRevB.1.910} {\bibfield  {journal} {\bibinfo
  {journal} {Phys. Rev. B}\ }\textbf {\bibinfo {volume} {1}},\ \bibinfo {pages}
  {910} (\bibinfo {year} {1970})}\BibitemShut {NoStop}%
\bibitem [{\citenamefont {Qiu}\ \emph {et~al.}(2013)\citenamefont {Qiu},
  \citenamefont {da~Jornada},\ and\ \citenamefont {Louie}}]{Qiu_2013}%
  \BibitemOpen
  \bibfield  {author} {\bibinfo {author} {\bibfnamefont {D.~Y.}\ \bibnamefont
  {Qiu}}, \bibinfo {author} {\bibfnamefont {F.~H.}\ \bibnamefont {da~Jornada}},
  \ and\ \bibinfo {author} {\bibfnamefont {S.~G.}\ \bibnamefont {Louie}},\
  }\href {\doibase 10.1103/PhysRevLett.111.216805} {\bibfield  {journal}
  {\bibinfo  {journal} {Phys. Rev. Lett.}\ }\textbf {\bibinfo {volume} {111}},\
  \bibinfo {pages} {216805} (\bibinfo {year} {2013})}\BibitemShut {NoStop}%
\bibitem [{\citenamefont {Molina-S\'anchez}\ \emph {et~al.}(2013)\citenamefont
  {Molina-S\'anchez}, \citenamefont {Sangalli}, \citenamefont {Hummer},
  \citenamefont {Marini},\ and\ \citenamefont {Wirtz}}]{Molina_2013}%
  \BibitemOpen
  \bibfield  {author} {\bibinfo {author} {\bibfnamefont {A.}~\bibnamefont
  {Molina-S\'anchez}}, \bibinfo {author} {\bibfnamefont {D.}~\bibnamefont
  {Sangalli}}, \bibinfo {author} {\bibfnamefont {K.}~\bibnamefont {Hummer}},
  \bibinfo {author} {\bibfnamefont {A.}~\bibnamefont {Marini}}, \ and\ \bibinfo
  {author} {\bibfnamefont {L.}~\bibnamefont {Wirtz}},\ }\href {\doibase
  10.1103/PhysRevB.88.045412} {\bibfield  {journal} {\bibinfo  {journal} {Phys.
  Rev. B}\ }\textbf {\bibinfo {volume} {88}},\ \bibinfo {pages} {045412}
  (\bibinfo {year} {2013})}\BibitemShut {NoStop}%
\bibitem [{\citenamefont {Zhu}\ \emph {et~al.}(2011)\citenamefont {Zhu},
  \citenamefont {Cheng},\ and\ \citenamefont {Schwingenschl\"ogl}}]{Zhu_2011}%
  \BibitemOpen
  \bibfield  {author} {\bibinfo {author} {\bibfnamefont {Z.~Y.}\ \bibnamefont
  {Zhu}}, \bibinfo {author} {\bibfnamefont {Y.~C.}\ \bibnamefont {Cheng}}, \
  and\ \bibinfo {author} {\bibfnamefont {U.}~\bibnamefont
  {Schwingenschl\"ogl}},\ }\href {\doibase 10.1103/PhysRevB.84.153402}
  {\bibfield  {journal} {\bibinfo  {journal} {Phys. Rev. B}\ }\textbf {\bibinfo
  {volume} {84}},\ \bibinfo {pages} {153402} (\bibinfo {year}
  {2011})}\BibitemShut {NoStop}%
\bibitem [{\citenamefont {Ferreira}\ \emph {et~al.}(2019)\citenamefont
  {Ferreira}, \citenamefont {Chaves}, \citenamefont {Peres},\ and\
  \citenamefont {Ribeiro}}]{Ferreira_2019}%
  \BibitemOpen
  \bibfield  {author} {\bibinfo {author} {\bibfnamefont {F.}~\bibnamefont
  {Ferreira}}, \bibinfo {author} {\bibfnamefont {A.~J.}\ \bibnamefont
  {Chaves}}, \bibinfo {author} {\bibfnamefont {N.~M.~R.}\ \bibnamefont
  {Peres}}, \ and\ \bibinfo {author} {\bibfnamefont {R.~M.}\ \bibnamefont
  {Ribeiro}},\ }\href {\doibase 10.1364/JOSAB.36.000674} {\bibfield  {journal}
  {\bibinfo  {journal} {J. Opt. Soc. Am. B}\ }\textbf {\bibinfo {volume}
  {36}},\ \bibinfo {pages} {674} (\bibinfo {year} {2019})}\BibitemShut
  {NoStop}%
\bibitem [{\citenamefont {Cudazzo}\ \emph {et~al.}(2016)\citenamefont
  {Cudazzo}, \citenamefont {Sponza}, \citenamefont {Giorgetti}, \citenamefont
  {Reining}, \citenamefont {Sottile},\ and\ \citenamefont
  {Gatti}}]{Cudazzo_2016}%
  \BibitemOpen
  \bibfield  {author} {\bibinfo {author} {\bibfnamefont {P.}~\bibnamefont
  {Cudazzo}}, \bibinfo {author} {\bibfnamefont {L.}~\bibnamefont {Sponza}},
  \bibinfo {author} {\bibfnamefont {C.}~\bibnamefont {Giorgetti}}, \bibinfo
  {author} {\bibfnamefont {L.}~\bibnamefont {Reining}}, \bibinfo {author}
  {\bibfnamefont {F.}~\bibnamefont {Sottile}}, \ and\ \bibinfo {author}
  {\bibfnamefont {M.}~\bibnamefont {Gatti}},\ }\href {\doibase
  10.1103/PhysRevLett.116.066803} {\bibfield  {journal} {\bibinfo  {journal}
  {Phys. Rev. Lett.}\ }\textbf {\bibinfo {volume} {116}},\ \bibinfo {pages}
  {066803} (\bibinfo {year} {2016})}\BibitemShut {NoStop}%
\bibitem [{\citenamefont {Galvani}\ \emph {et~al.}(2016)\citenamefont
  {Galvani}, \citenamefont {Paleari}, \citenamefont {Miranda}, \citenamefont
  {Molina-S\'anchez}, \citenamefont {Wirtz}, \citenamefont {Latil},
  \citenamefont {Amara},\ and\ \citenamefont {Ducastelle}}]{Galvani_2016}%
  \BibitemOpen
  \bibfield  {author} {\bibinfo {author} {\bibfnamefont {T.}~\bibnamefont
  {Galvani}}, \bibinfo {author} {\bibfnamefont {F.}~\bibnamefont {Paleari}},
  \bibinfo {author} {\bibfnamefont {H.~P.~C.}\ \bibnamefont {Miranda}},
  \bibinfo {author} {\bibfnamefont {A.}~\bibnamefont {Molina-S\'anchez}},
  \bibinfo {author} {\bibfnamefont {L.}~\bibnamefont {Wirtz}}, \bibinfo
  {author} {\bibfnamefont {S.}~\bibnamefont {Latil}}, \bibinfo {author}
  {\bibfnamefont {H.}~\bibnamefont {Amara}}, \ and\ \bibinfo {author}
  {\bibfnamefont {F.~m.~c.}\ \bibnamefont {Ducastelle}},\ }\href {\doibase
  10.1103/PhysRevB.94.125303} {\bibfield  {journal} {\bibinfo  {journal} {Phys.
  Rev. B}\ }\textbf {\bibinfo {volume} {94}},\ \bibinfo {pages} {125303}
  (\bibinfo {year} {2016})}\BibitemShut {NoStop}%
\bibitem [{\citenamefont {Berseneva}\ \emph {et~al.}(2013)\citenamefont
  {Berseneva}, \citenamefont {Gulans}, \citenamefont {Krasheninnikov},\ and\
  \citenamefont {Nieminen}}]{Berseneva_2013}%
  \BibitemOpen
  \bibfield  {author} {\bibinfo {author} {\bibfnamefont {N.}~\bibnamefont
  {Berseneva}}, \bibinfo {author} {\bibfnamefont {A.}~\bibnamefont {Gulans}},
  \bibinfo {author} {\bibfnamefont {A.~V.}\ \bibnamefont {Krasheninnikov}}, \
  and\ \bibinfo {author} {\bibfnamefont {R.~M.}\ \bibnamefont {Nieminen}},\
  }\href {\doibase 10.1103/PhysRevB.87.035404} {\bibfield  {journal} {\bibinfo
  {journal} {Phys. Rev. B}\ }\textbf {\bibinfo {volume} {87}},\ \bibinfo
  {pages} {035404} (\bibinfo {year} {2013})}\BibitemShut {NoStop}%
\bibitem [{\citenamefont {\ifmmode~\mbox{\c{S}}\else \c{S}\fi{}ahin}\ \emph
  {et~al.}(2009)\citenamefont {\ifmmode~\mbox{\c{S}}\else \c{S}\fi{}ahin},
  \citenamefont {Cahangirov}, \citenamefont {Topsakal}, \citenamefont
  {Bekaroglu}, \citenamefont {Akturk}, \citenamefont {Senger},\ and\
  \citenamefont {Ciraci}}]{Ciraci_2009}%
  \BibitemOpen
  \bibfield  {author} {\bibinfo {author} {\bibfnamefont {H.}~\bibnamefont
  {\ifmmode~\mbox{\c{S}}\else \c{S}\fi{}ahin}}, \bibinfo {author}
  {\bibfnamefont {S.}~\bibnamefont {Cahangirov}}, \bibinfo {author}
  {\bibfnamefont {M.}~\bibnamefont {Topsakal}}, \bibinfo {author}
  {\bibfnamefont {E.}~\bibnamefont {Bekaroglu}}, \bibinfo {author}
  {\bibfnamefont {E.}~\bibnamefont {Akturk}}, \bibinfo {author} {\bibfnamefont
  {R.~T.}\ \bibnamefont {Senger}}, \ and\ \bibinfo {author} {\bibfnamefont
  {S.}~\bibnamefont {Ciraci}},\ }\href {\doibase 10.1103/PhysRevB.80.155453}
  {\bibfield  {journal} {\bibinfo  {journal} {Phys. Rev. B}\ }\textbf {\bibinfo
  {volume} {80}},\ \bibinfo {pages} {155453} (\bibinfo {year}
  {2009})}\BibitemShut {NoStop}%
\bibitem [{\citenamefont {Blase}\ \emph {et~al.}(1995)\citenamefont {Blase},
  \citenamefont {Rubio}, \citenamefont {Louie},\ and\ \citenamefont
  {Cohen}}]{Blase_1995}%
  \BibitemOpen
  \bibfield  {author} {\bibinfo {author} {\bibfnamefont {X.}~\bibnamefont
  {Blase}}, \bibinfo {author} {\bibfnamefont {A.}~\bibnamefont {Rubio}},
  \bibinfo {author} {\bibfnamefont {S.~G.}\ \bibnamefont {Louie}}, \ and\
  \bibinfo {author} {\bibfnamefont {M.~L.}\ \bibnamefont {Cohen}},\ }\href
  {\doibase 10.1103/PhysRevB.51.6868} {\bibfield  {journal} {\bibinfo
  {journal} {Phys. Rev. B}\ }\textbf {\bibinfo {volume} {51}},\ \bibinfo
  {pages} {6868} (\bibinfo {year} {1995})}\BibitemShut {NoStop}%
\bibitem [{\citenamefont {Wirtz}\ \emph {et~al.}(2006)\citenamefont {Wirtz},
  \citenamefont {Marini},\ and\ \citenamefont {Rubio}}]{Wirtz_2006}%
  \BibitemOpen
  \bibfield  {author} {\bibinfo {author} {\bibfnamefont {L.}~\bibnamefont
  {Wirtz}}, \bibinfo {author} {\bibfnamefont {A.}~\bibnamefont {Marini}}, \
  and\ \bibinfo {author} {\bibfnamefont {A.}~\bibnamefont {Rubio}},\ }\href
  {\doibase 10.1103/PhysRevLett.96.126104} {\bibfield  {journal} {\bibinfo
  {journal} {Phys. Rev. Lett.}\ }\textbf {\bibinfo {volume} {96}},\ \bibinfo
  {pages} {126104} (\bibinfo {year} {2006})}\BibitemShut {NoStop}%
\bibitem [{\citenamefont {Li}\ \emph {et~al.}(2017)\citenamefont {Li},
  \citenamefont {Kim}, \citenamefont {Jin}, \citenamefont {Ye}, \citenamefont
  {Qiu}, \citenamefont {da~Jornada}, \citenamefont {Shi}, \citenamefont {Chen},
  \citenamefont {Zhang}, \citenamefont {Yang}, \citenamefont {Watanabe},
  \citenamefont {Taniguchi}, \citenamefont {Ren}, \citenamefont {Louie},
  \citenamefont {Chen}, \citenamefont {Zhang},\ and\ \citenamefont
  {Wang}}]{Li_2017}%
  \BibitemOpen
  \bibfield  {author} {\bibinfo {author} {\bibfnamefont {L.}~\bibnamefont
  {Li}}, \bibinfo {author} {\bibfnamefont {J.}~\bibnamefont {Kim}}, \bibinfo
  {author} {\bibfnamefont {C.}~\bibnamefont {Jin}}, \bibinfo {author}
  {\bibfnamefont {G.~J.}\ \bibnamefont {Ye}}, \bibinfo {author} {\bibfnamefont
  {D.~Y.}\ \bibnamefont {Qiu}}, \bibinfo {author} {\bibfnamefont {F.~H.}\
  \bibnamefont {da~Jornada}}, \bibinfo {author} {\bibfnamefont
  {Z.}~\bibnamefont {Shi}}, \bibinfo {author} {\bibfnamefont {L.}~\bibnamefont
  {Chen}}, \bibinfo {author} {\bibfnamefont {Z.}~\bibnamefont {Zhang}},
  \bibinfo {author} {\bibfnamefont {F.}~\bibnamefont {Yang}}, \bibinfo {author}
  {\bibfnamefont {K.}~\bibnamefont {Watanabe}}, \bibinfo {author}
  {\bibfnamefont {T.}~\bibnamefont {Taniguchi}}, \bibinfo {author}
  {\bibfnamefont {W.}~\bibnamefont {Ren}}, \bibinfo {author} {\bibfnamefont
  {S.~G.}\ \bibnamefont {Louie}}, \bibinfo {author} {\bibfnamefont {X.~H.}\
  \bibnamefont {Chen}}, \bibinfo {author} {\bibfnamefont {Y.}~\bibnamefont
  {Zhang}}, \ and\ \bibinfo {author} {\bibfnamefont {F.}~\bibnamefont {Wang}},\
  }\href {\doibase 10.1038/nnano.2016.171} {\bibfield  {journal} {\bibinfo
  {journal} {Nature Nanotech.}\ }\textbf {\bibinfo {volume} {12}},\ \bibinfo
  {pages} {21} (\bibinfo {year} {2017})}\BibitemShut {NoStop}%
\bibitem [{\citenamefont {Yoon}\ \emph {et~al.}(2021)\citenamefont {Yoon},
  \citenamefont {Kim}, \citenamefont {Seo}, \citenamefont {Shin}, \citenamefont
  {Song}, \citenamefont {Kim}, \citenamefont {Watanabe}, \citenamefont
  {Taniguchi}, \citenamefont {Lee}, \citenamefont {Jo},\ and\ \citenamefont
  {et~al.}}]{Yoon_2021}%
  \BibitemOpen
  \bibfield  {author} {\bibinfo {author} {\bibfnamefont {S.}~\bibnamefont
  {Yoon}}, \bibinfo {author} {\bibfnamefont {T.}~\bibnamefont {Kim}}, \bibinfo
  {author} {\bibfnamefont {S.-Y.}\ \bibnamefont {Seo}}, \bibinfo {author}
  {\bibfnamefont {S.-H.}\ \bibnamefont {Shin}}, \bibinfo {author}
  {\bibfnamefont {S.-B.}\ \bibnamefont {Song}}, \bibinfo {author}
  {\bibfnamefont {B.~J.}\ \bibnamefont {Kim}}, \bibinfo {author} {\bibfnamefont
  {K.}~\bibnamefont {Watanabe}}, \bibinfo {author} {\bibfnamefont
  {T.}~\bibnamefont {Taniguchi}}, \bibinfo {author} {\bibfnamefont {G.-H.}\
  \bibnamefont {Lee}}, \bibinfo {author} {\bibfnamefont {M.-H.}\ \bibnamefont
  {Jo}}, \ and\ \bibinfo {author} {\bibnamefont {et~al.}},\ }\href {\doibase
  10.1103/physrevb.103.l041407} {\bibfield  {journal} {\bibinfo  {journal}
  {Phys. Rev. B}\ }\textbf {\bibinfo {volume} {103}} (\bibinfo {year} {2021}),\
  10.1103/physrevb.103.l041407}\BibitemShut {NoStop}%
\bibitem [{\citenamefont {Giannozzi}\ \emph {et~al.}(2020)\citenamefont
  {Giannozzi}, \citenamefont {Baseggio}, \citenamefont {Bonfà}, \citenamefont
  {Brunato}, \citenamefont {Car}, \citenamefont {Carnimeo}, \citenamefont
  {Cavazzoni}, \citenamefont {de~Gironcoli}, \citenamefont {Delugas},
  \citenamefont {Ferrari~Ruffino}, \citenamefont {Ferretti}, \citenamefont
  {Marzari}, \citenamefont {Timrov}, \citenamefont {Urru},\ and\ \citenamefont
  {Baroni}}]{QE_2020}%
  \BibitemOpen
  \bibfield  {author} {\bibinfo {author} {\bibfnamefont {P.}~\bibnamefont
  {Giannozzi}}, \bibinfo {author} {\bibfnamefont {O.}~\bibnamefont {Baseggio}},
  \bibinfo {author} {\bibfnamefont {P.}~\bibnamefont {Bonfà}}, \bibinfo
  {author} {\bibfnamefont {D.}~\bibnamefont {Brunato}}, \bibinfo {author}
  {\bibfnamefont {R.}~\bibnamefont {Car}}, \bibinfo {author} {\bibfnamefont
  {I.}~\bibnamefont {Carnimeo}}, \bibinfo {author} {\bibfnamefont
  {C.}~\bibnamefont {Cavazzoni}}, \bibinfo {author} {\bibfnamefont
  {S.}~\bibnamefont {de~Gironcoli}}, \bibinfo {author} {\bibfnamefont
  {P.}~\bibnamefont {Delugas}}, \bibinfo {author} {\bibfnamefont
  {F.}~\bibnamefont {Ferrari~Ruffino}}, \bibinfo {author} {\bibfnamefont
  {A.}~\bibnamefont {Ferretti}}, \bibinfo {author} {\bibfnamefont
  {N.}~\bibnamefont {Marzari}}, \bibinfo {author} {\bibfnamefont
  {I.}~\bibnamefont {Timrov}}, \bibinfo {author} {\bibfnamefont
  {A.}~\bibnamefont {Urru}}, \ and\ \bibinfo {author} {\bibfnamefont
  {S.}~\bibnamefont {Baroni}},\ }\href {\doibase 10.1063/5.0005082} {\bibfield
  {journal} {\bibinfo  {journal} {J. Chem. Phys.}\ }\textbf {\bibinfo {volume}
  {152}},\ \bibinfo {pages} {154105} (\bibinfo {year} {2020})}\BibitemShut
  {NoStop}%
\bibitem [{\citenamefont {Perdew}\ \emph {et~al.}(1996)\citenamefont {Perdew},
  \citenamefont {Burke},\ and\ \citenamefont {Ernzerhof}}]{PBE}%
  \BibitemOpen
  \bibfield  {author} {\bibinfo {author} {\bibfnamefont {J.~P.}\ \bibnamefont
  {Perdew}}, \bibinfo {author} {\bibfnamefont {K.}~\bibnamefont {Burke}}, \
  and\ \bibinfo {author} {\bibfnamefont {M.}~\bibnamefont {Ernzerhof}},\ }\href
  {\doibase 10.1103/PhysRevLett.77.3865} {\bibfield  {journal} {\bibinfo
  {journal} {Phys. Rev. Lett.}\ }\textbf {\bibinfo {volume} {77}},\ \bibinfo
  {pages} {3865} (\bibinfo {year} {1996})}\BibitemShut {NoStop}%
\bibitem [{\citenamefont {Bruneval}\ and\ \citenamefont
  {Gonze}(2008)}]{Bruneval_2008}%
  \BibitemOpen
  \bibfield  {author} {\bibinfo {author} {\bibfnamefont {F.}~\bibnamefont
  {Bruneval}}\ and\ \bibinfo {author} {\bibfnamefont {X.}~\bibnamefont
  {Gonze}},\ }\href {\doibase 10.1103/PhysRevB.78.085125} {\bibfield  {journal}
  {\bibinfo  {journal} {Phys. Rev. B}\ }\textbf {\bibinfo {volume} {78}},\
  \bibinfo {pages} {085125} (\bibinfo {year} {2008})}\BibitemShut {NoStop}%
\bibitem [{\citenamefont {Martin}\ \emph {et~al.}(2016)\citenamefont {Martin},
  \citenamefont {Reining},\ and\ \citenamefont
  {Ceperley}}]{martin_reining_ceperley_2016}%
  \BibitemOpen
  \bibfield  {author} {\bibinfo {author} {\bibfnamefont {R.~M.}\ \bibnamefont
  {Martin}}, \bibinfo {author} {\bibfnamefont {L.}~\bibnamefont {Reining}}, \
  and\ \bibinfo {author} {\bibfnamefont {D.~M.}\ \bibnamefont {Ceperley}},\
  }\href {\doibase 10.1017/CBO9781139050807} {\emph {\bibinfo {title}
  {Interacting Electrons: Theory and Computational Approaches}}}\ (\bibinfo
  {publisher} {Cambridge University Press},\ \bibinfo {year}
  {2016})\BibitemShut {NoStop}%
\end{thebibliography}%


%\printbibliography
%
\end{document}